\documentclass [prb,twocolumn,notitlepage,superscriptaddress,eqsecnum,floatfix] {revtex4-1}
\usepackage{amsmath}
\usepackage{graphicx}
\usepackage{lmodern}
\usepackage{amsmath}
\usepackage{bm}

\usepackage{color}
\usepackage{amssymb}
\newcommand{\beq} {\begin{equation}}
\newcommand{\eeq} {\end{equation}}
\newcommand{\bea} {\begin{eqnarray}}
\newcommand{\eea} {\end{eqnarray}}
\newcommand{\be} {\begin{equation}}
\newcommand{\ee} {\end{equation}}
\renewcommand{\(}{\left(}
\renewcommand{\)}{\right)}
\renewcommand{\[}{\left[}
\renewcommand{\]}{\right]}

\DeclareMathOperator{\sgn}{sgn}
\DeclareMathOperator{\Tr}{Tr}

\newcommand{\U} {{\mathrm {U(1)}}}

\renewcommand{\k} {{\bm k}}

\begin{document}

\title {Topological superconducting phases from inversion symmetry breaking order in spin-orbit-coupled systems }
\author{Yuxuan Wang}
\affiliation{Department of Physics and Institute of Condensed Matter Theory, University of Illinois at Urbana-Champaign, 1110 West Green Street, Urbana, Illinois 61801-3080, USA }
\author{Gil Young Cho}
\affiliation{Department of Physics, Korea Advanced Institute of Science and Technology, Daejeon 305-701, Korea}
\author{Taylor L. Hughes}
\affiliation{Department of Physics and Institute of Condensed Matter Theory, University of Illinois at Urbana-Champaign, 1110 West Green Street, Urbana, Illinois 61801-3080, USA }
\author{Eduardo Fradkin}
\affiliation{Department of Physics and Institute of Condensed Matter Theory, University of Illinois at Urbana-Champaign, 1110 West Green Street, Urbana, Illinois 61801-3080, USA }
\date{\today}

\begin{abstract}
We analyze the superconducting instabilities in the vicinity of the quantum-critical point of an inversion symmetry breaking order. We first show that the fluctuations of the inversion symmetry breaking order lead to two degenerate superconducting (SC) instabilities, one  in the $s$-wave channel, and the other in a time-reversal invariant odd-parity pairing channel (the simplest case being the same as the of $^3$He-B phase). {Remarkably, we find that unlike many well-known examples, the selection of the pairing symmetry of the condensate is independent of the momentum-space structure of the collective mode that mediates the pairing interaction.}
We found that this degeneracy is a {result of the existence of a conserved fermionic helicity}, $\chi$, and the two degenerate channels correspond to even and odd combinations of SC order parameters 
{with} $\chi=\pm1$. As a result, the system has an enlarged symmetry $\U\times\U$, {with} each $\U$ corresponding to one value of {the helicity} $\chi$. 
{Because of the enlarged symmetry, this system admits exotic topological defects such as a fractional quantum vortex, which we show has a Majorana zero mode bound at its core.}
We discuss how the enlarged symmetry can be lifted by small perturbations, such as the Coulomb interaction or Fermi surface splitting in the presence of broken inversion symmetry, and we show that the resulting superconducting state can be topological or trivial depending on parameters. {The $\U\times\U$ symmetry is restored at the phase boundary between the topological and trivial SC states, and allows for a  transition between topologically distinct SC phases without the vanishing of the order parameter.} We present a global phase diagram of the superconducting states and discuss possible experimental implications.
\end{abstract}
\maketitle

\section{Introduction}

Over the last few decades, it has become clear the classification of states of matter extends beyond the paradigm of Landau's spontaneous symmetry-breaking.\cite{wenbook,Qi-2010,Hasan-2010} Even states that have the same symmetries can display distinct properties due to the topology embedded in their many-body wavefunctions. For example, three-dimensional (3D) superconducting (SC) states, which all break $\U$ gauge invariance, but maintain time-reversal symmetry, can be further classified by their topological properties.\cite{qihughesraghuzhang,Ryu-2008} Different from conventional $s$-wave superconductors, these so-called topological superconductors exhibit exotic quasiparticle excitations at their boundaries, and at the cores of various (vortex) defects. Non-trivial topological superconductors have exciting potential applications in, e.g., topological quantum computation,\cite{Nayak-2008} and this has sparked intense theoretical and experimental efforts in search of these materials.\cite{Fu-Berg-2010,Kriener-2011,Levy-2013,Fu-2014,Wan-2014,Nikosai-2012,Scheurer-2015,Hosur-2014,Yuan-2014,Yoshida-2015,Yang-2015,Ando-2015} Unlike {most} topological insulators, whose topological properties stem from their band structure alone, topological superconductors require cooperation between band structure and interactions for their existence. This makes their prediction in real materials more challenging. Indeed, most of the predicted topological superconductors are unconventional superconductors, and the pairing symmetry is usually odd-parity, such as $p$-wave or $f$-wave.\cite{Fu-Berg-2010,Sato-2010}

In the search for unconventional superconductivity, one important 
scenario
is where Cooper pairing is mediated by the soft bosonic fluctuations of collective modes (sometimes in the vicinity of a quantum-critical point (QCP)), with examples ranging from liquid $^3$He\cite{Leggett-1975,Vollhardt-1990} to heavy-fermion materials.\cite{Varma-1986} This scenario has {often} been suggested as a possible mechanism for the cuprates~\cite{Scalapino-1995,Abanov-2003,Metlitski-2010,Wang-2014} and Fe-pnictide~\cite{Chubukov-2012} superconductors. In this picture, the QCP underlying the superconducting phase plays an important role in renormalizing the normal (non-superconducting) state properties, and equally importantly helps determine the pairing symmetry of the superconducting state.

Along this line of reasoning, {Kozii and Fu\cite{Fu-2015} analyzed} the superconducting instabilities mediated by the fluctuations of an order that is odd under inversion symmetry and invariant under time-reversal, presumably close to a QCP where the inversion symmetry is spontaneously broken.
Such an order parameter can emerge from spin-orbit coupled systems with spontaneously broken inversion symmetry,\cite{Fu-2015a,Wu-2004,Wu-2007} and will couple to the {fermionic degrees of freedom}  via a term of the form $\sim \phi c^\dagger_{\k\alpha}({\bm d_\k}\cdot{\bm \sigma}_{\alpha\beta}) c_{\k\beta}$, where ${\bm \sigma}$ transforms as spin under inversion and time-reversal, and ${\bm d_\k}$ is an odd function of $\k$.
 In the ordered state, the Fermi surface splits into two, {as a result of } the broken inversion symmetry. {On the other hand, in} the disordered phase, the fluctuations of the inversion-breaking order mediate an effective four-fermion interaction which is attractive in the Cooper channel.
 As a result, it was found that, together with conventional $s$-wave pairing, there exists {also} a time-reversal invariant odd-parity pairing instability. In the simplest case, where ${\bm d}_\k\propto \k$, it is a $p$-wave pairing, the same as in the superfluid  $^3$He-B phase. In that work the authors worked under a restricted, ``zeroth order", approximation where the pairing interaction mediated by the parity fluctuations is assumed to be independent of the transferred momentum and frequency. Within this treatment, it was found that, remarkably, the two superconducting channels have the same onset temperature $T_c$, and it was speculated that other interactions, such as the Coulomb interaction, or an external Zeeman field, could lift this degeneracy and favor the odd-parity superconducting state. Furthermore, it was also proposed that the pyrochlore oxide Cd$_2$Re$_2$O$_7$ and doped SrTiO$_3$ heterostructures are candidate materials for this type of unconventional superconductor. 

Unfortunately this analysis leaves open a number of important issues. First, the analysis was based on approximating {the  propagator mediated by parity fluctuations} to be independent of momentum and frequency, while in reality it becomes strongly dependent on the transferred momentum and frequency close to the onset of the inversion symmetry breaking order. Since the momentum dependence of the effective interaction usually plays an important role in determining the pairing symmetry,  an immediate open question is how it affects the relative strength of the odd-parity pairing instability compared with the conventional $s$-wave pairing when treated with a more realistic analysis. 

Second, the degeneracy between $s$-wave and odd-parity pairing channels indicates that the two are strong competitors below $T_c$, and their interplay in the superconducting phase remains to be addressed. It is possible that either one may order, or they may even coexist. Indeed,  if they do coexist, there is also the question of whether their relative phase becomes locked, which could lead to a spontaneous breaking of some discrete symmetry. To study their interplay, we will explicitly expand the free energy to quartic order in terms of the SC order parameters and determine the coefficients. Third, the topological properties of the superconducting state(s) remain to be identified and analyzed, particularly for the possible coexistence states of $s$-wave and odd-parity SC orders.
 
In this article we plan to address all of these issues. Throughout we will usually use the simplest case of the inversion-breaking order as an example, i.e., where ${\bm d}_\k\propto \k$ and the resultant odd-parity pairing is $p$-wave. We will explicitly show that 
 the $s$-wave and the $p$-wave channels are degenerate, \emph{even} when an arbitrary momentum and frequency dependence of the effective interaction is included. We find that this seemingly accidental degeneracy has a deeper reason, namely, the effective interaction conserves the helicity ${\bm \sigma}\cdot\k$ of fermions on the FS, 
 since the Yukawa coupling between low-energy fermions of opposite helicity and the mediating bosonic mode vanishes.
As a result, the fermions of the two different helicities pair independently, leading to two independent superconducting order parameters $\Delta_1$ and $\Delta_2$. The even and odd combinations of $\Delta_1$ and $\Delta_2$ are exactly the $s$-wave pairing and the odd-parity pairing, namely, $\Delta_{s/p}\sim\Delta_1\pm\Delta_2$. What is more, from this reasoning it is clear that the system has a $\U\times\U$ symmetry at this level. 
%

%
In a realistic system, various small perturbations can lift the degeneracy between $s$-wave and odd-parity pairing channels. We show that  the Coulomb interaction lifts the $\U\times\U$ symmetry, and favors the odd-parity pairing; heuristically because it avoids the short-range repulsion. 
On the other hand, the FS splitting in the broken inversion symmetry phase also lifts this enlarged symmetry because the FS mismatch gives a ``residual" coupling between fermions on FS's of opposite helicity. This residual interaction favors the $s$-wave pairing instead. The combination of these two effects leads to, in the simplest case where ${\bm d}_\k\propto \k$, a $p+s$ state, and  which 
superconducting component, 
$\Delta_p$ or $\Delta_s,$ is larger depends on which effect is stronger. 
We identify that the latter effect is stronger as the system goes deeper into the ordered state, for which the FS splitting is larger, and present a global phase diagram for the SC orders in Fig.\ \ref{phase} as a function of temperature and an extra parameter that tunes the inversion breaking order.

Interestingly, we will show that if $\vert \Delta_p \vert > \vert\Delta_s\vert$ then the system is a topological superconductor, and if  $\vert \Delta_p \vert  < \vert\Delta_s\vert$  then it is a topologically trivial one.
At the phase boundary between the topological and trivial phases, {we show that, at higher temperatures, the transition between topologically distinct states occurs through the vanishing of the SC order parameter on one or more FS's, while at lower temperatures, the transition circumvents the vanishing of the SC order parameter by taking a path that effectively breaks time-reversal symmetry due to strong fluctuations of the relative superconducting phase.\cite{Nagaosa-2013}}
Since the two effects that compete to choose the $s$- and $p$-wave pairing are tuned to cancel at this phase boundary,  the $\U\times\U$ symmetry of the free energy is restored. Such an enlarged symmetry enables exotic vortex defects, including a fractional quantum vortex\cite{babaev,Berg-2009}. The core of the half quantum vortex traps a Majorana zero mode and has non-Abelian braiding statistics{\cite{readgreen,ivanov}}.
  
 The remainder of this paper is organized as follows. In Sec. II we introduce the model for the pairing problem near the onset of an inversion symmetry breaking order. In Sec. III we show the $s$-wave channel and the odd-parity channels are degenerate. In Secs. IV  and V we analyze the origin of this degeneracy, derive the Ginzburg-Landau free energy up to quartic order, and present the global phase diagram for the superconducting order. In Sec. VI we discuss the topological properties of the superconducting phases and other relations to experiments. Finally, we conclude in Sec. VII and present extra calculational details in Appendix A.

\section{Model}

We consider an isotropic, itinerant electronic system in the vicinity of an inversion symmetry breaking QCP. 
Generically, the inversion symmetry breaking order {parameter}  couples with the fermion via a bilinear form~\cite{Fu-2015}
\begin{align}
\hat Q=\sum_{{\bm k},\alpha\beta}\Gamma_{\alpha\beta}({\bm k}) c^\dagger_{{\bm k}\alpha} c_{{\bm k}\beta}
\end{align}
where $\Gamma^\dagger(k)=\Gamma(k)$, and
\begin{align}
\Gamma({k})=\langle\phi\rangle{\bm d}_{\bm k} \cdot {\bm \sigma},
\end{align}
where {the ``director''} ${\bm d}_{\bm k}=-{\bm d}_{\bm -k}$ is odd under inversion, and the bosonic field $\phi$ transforms as a singlet under inversion and time reversal.
The order parameter $\Gamma$ breaks inversion symmetry but not time-reversal symmetry, which dictates that the Pauli matrices ${\bm \sigma}=(\sigma^1,\sigma^2,\sigma^3)$ are even under inversion and odd under time-reversal. They can represent the actual electron spin operators~\cite{Wu-2004} (with the caveat that in order to stabilize the inversion breaking order in a $p$-wave channel a small but finite intrinsic spin-orbit coupling is required~\cite{Wu-2007}), or pseudospin operators in the so-called ``manifestly covariant Bloch basis"~\cite{Fu-2015a} (MCBB) extracted from a multiband spin-orbit-coupled system.
 In the ordered phase, $\langle \phi \rangle$ condenses and   splits of the initially doubly-degenerate Fermi surface (FS). Importantly, on each piece of the FS, {the (pseudo)spin}  $\bm \sigma$ is aligned or anti-aligned with ${\bm d}_{\bm k}$.

In the disordered phase, $\langle\phi\rangle=0$, but close to the onset of the inversion symmetry breaking order {phase}, the fluctuations of $\phi=\phi(q)$ are soft and give rise to an effective four-fermion interaction of the form 
\begin{align} 
S_V=&-\sum_{q}V(q) \bar Q({\bm q})  \bar Q({-\bm q})
\label{V}
\end{align}
where $q=(\omega,{\bm q})$. Here $V(q)\equiv \langle \phi(q)\phi(-q)\rangle$ is the bosonic propagator of the order parameter field $\phi$ associated with the inversion symmetry breaking, and it generally depends on both the momentum and frequency transfer (and is what was approximated to be independent of both in Ref.\ \onlinecite{Fu-2015}).  The boson-fermion vertex is given by a Yukawa-type coupling term
\begin{align}
\bar Q({\bm q})
=&\frac12\sum_{\bm k} c^\dagger_{{\bm{k+q}} \alpha}[({\bm d}_{\bm k}+{\bm d}_{{\bm{k+q}}}) \cdot {\bm\sigma}_{\alpha\beta}]c_{\bm k\beta}.
\label{eq4}
\end{align}
In the following sections we consider the types of superconducting order mediated by this effective interaction.

\section{Degeneracy of $s$-wave and odd-parity superconducting channels}
In this Section, we study the superconducting instabilities from the interaction given by Eq.\ (\ref{V}). 
We first consider a
 two-dimensional (2D) system with an isotropic (circular) Fermi surface (FS) with a parabolic fermionic dispersion. At the end of this Section we generalize this {treatment} to the isotropic 3D case with a spherical FS. 
We further assume a specific case where ${\bm d}_\k=\lambda\k/k_F$, where $k_F$ is the Fermi momentum and $\lambda$ is a dimensionless parameter. We will show that in this case the odd-parity superconducting pairing is naturally of $p$-wave symmetry, and we will later generalize the results to other forms of $\bm d_\k$. 
 The two superconducting order parameters in the $s$-wave channel and $p$-wave channel that we consider are given by
\begin{align}
{\mathcal H}_{s}=&\Delta^s_{\alpha\beta}c^\dagger_{\k\alpha}c^\dagger_{-\k\beta}=\Delta^s\; i\sigma^y_{\alpha\beta} \; c^\dagger_{\k\alpha}c^\dagger_{-\k\beta},~{\rm and} \nonumber\\
{\mathcal H}_{p}=&\Delta^p_{\alpha\beta}c^\dagger_{\k\alpha}c^\dagger_{-\k\beta}=\Delta^p \; \hat\k\cdot \; (i\bm\sigma \sigma^y)_{\alpha\beta}\; c^\dagger_{\k\alpha}c^\dagger_{-\k\beta}.
\label{sp}
\end{align}
In particular, for $\hat\k=(k_x,k_y)/k_F$, the $p$-wave order parameter is identical to {that of} the $^3$He-B phase,\cite{Leggett-1975,Vollhardt-1990} which is odd in parity, but is invariant under time-reversal and rotation in both spin and momentum space. It is well-known that this form of $p$-wave pairing generates a topological superconducting phase.\cite{qihughesraghuzhang,Ryu-2008} For simplicity, we will refer to this specific type of order as $p$-wave in the following.
\begin{figure}
\includegraphics[width=\columnwidth]{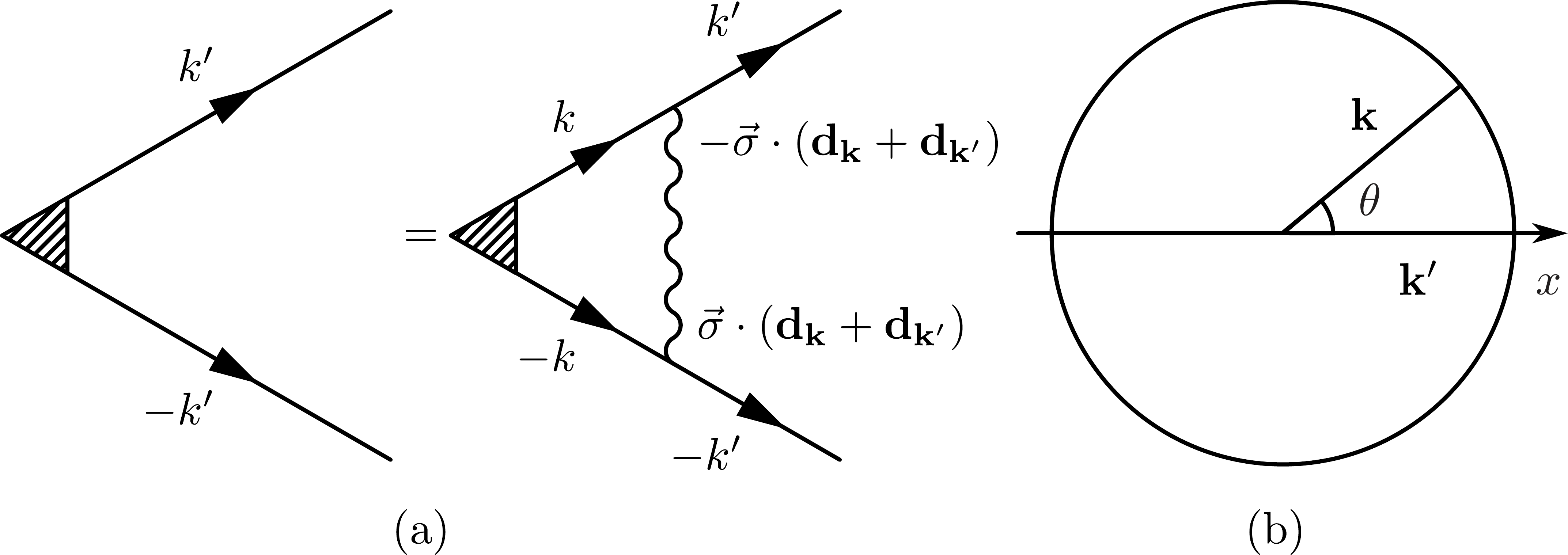}
\caption{(a) The diagrams for the linear gap equations for the SC orders $\Delta^s$ and $\Delta^p$, represented by the shaded triangle. (b) The fermion momenta $\bm k$ and $\bm k'$ on the FS in the rotated basis [see Eq.\ (\ref{sp2})].}
\label{fig:gap}
\end{figure}

Now let us consider the weak coupling case, for which the system is close to, but still away, from {the} quantum critical point (QCP) of  the inversion-breaking order. In this situation, contributions to the bosonic and fermionic self-energies can be neglected {at} the lowest-order {of} approximation, and the SC instability {is obtained by}  summing up {a suitable set of} ladder-type diagrams, which we show in a compact form in Fig.\ \ref{fig:gap}(a).

The {linearized} gap equations, that determine the ordering instabilities, can be expressed as
\begin{align}
\Delta_{\alpha\beta}(k')&=\frac14\sum_{k}G(k)G(-k)V(k-k') \Delta_{\delta\gamma}(k) \nonumber\\
&\times [({\bm d}_{\bm k}+{\bm d}_{\bm k'}) \cdot {\bm\sigma}_{\alpha\delta}][({\bm d}_{-\bm k}+{\bm d}_{-\bm k'}) \cdot {\bm\sigma}_{\beta\gamma}]
\label{eq7}
\end{align}
where $\Delta$ is either $\Delta^s$ or $\Delta^p$. {Here }
\begin{equation}
G(k)=G(\omega_m,\k)=\frac{1}{i\omega_m-{\bm v}_F\cdot \k}
\end{equation}
 is the fermionic Green function near the FS, and $V(k-k')$ is the propagator of the bosonic field $\phi$,
\begin{align}
V(k-k')=\frac{1}{(\omega_m-\omega_m')^2+(\k-\k')^2+\xi^{-2}}
\end{align}
where $\xi$ is the correlation length of the fluctuations of the inversion symmetry breaking order {parameter}.

The product of Green functions in Eq.\ \eqref{eq7} is logarithmically divergent in the infrared, and 
in the limit $T\ll v_F\xi^{-1}$, the integration over momentum perpendicular to the FS sets both $\k$ and $\k'$ on the FS (i.e., the Eliashberg approximation). For a circular FS, and for ${\bm d_\k}=\lambda\hat \k$, the SC gap becomes a function of FS angle only, $\Delta(\k)=\Delta(\hat \k)$. After integrating over the perpendicular momentum, we then have
\begin{align} 
\Delta_{\alpha\beta}(\hat \k')=&-\frac{N(0)\lambda^2}{4}\log\Big(\frac{\Lambda}{T}\Big) \int\frac{d\theta(\hat\k)}{2\pi}{V(\hat \k-\hat\k')} \Delta_{\delta\gamma}(\hat \k)\nonumber\\
&\times\[ (\hat \k+\hat \k')\cdot \bm\sigma_{\alpha\delta}\] \[ (\hat \k+\hat \k')\cdot \bm\sigma_{\beta\gamma}\],
\label{gap}
\end{align}
where $\theta(\hat\k)$ is the FS angle, $\Lambda$ is the upper energy cutoff, and $N(0)$ is the density of states near the FS. For convenience we define \begin{align}
\kappa_c\equiv\frac{N(0)\lambda^2}{4}\log\frac{\Lambda}{T}.
\label{eqA}
\end{align}

Next we show that both $\Delta^s$ and $\Delta^p$ are eigenfunctions of the angular integration kernel in Eq.\  (\ref{gap}), and they are degenerate. To see this, it is most convenient to work in a rotated basis where $\hat\k'=\hat {\bm x}$, and we define the angle between $\k$ and $\k'$ as $\theta$ [see Fig.\ \ref{fig:gap}(b)]. In this basis, 
\begin{align}
(\hat \k+\hat \k')\cdot \bm\sigma_{\alpha\delta}&=\sigma^x_{\alpha\delta}(1+\cos\theta)+\sigma^y_{\alpha\delta}\sin\theta, \nonumber\\
\Delta^s_{\delta\gamma}(\hat \k)&=\Delta^s(i\sigma^y)_{\delta\gamma}, \nonumber\\
\Delta^p_{\delta\gamma}(\hat \k)&=\Delta^p(-\cos\theta\sigma^z_{\delta\gamma}+i\sin\theta\delta_{\delta\gamma}).
\label{sp2}
\end{align}
Plugging these identities  into the right hand side of Eq.\ (\ref{gap}), we find that the right hand side for $\Delta^{s,p}$ is given by
\begin{align}
I^s=&-\kappa_c\Delta^s\int \frac{d\theta}{2\pi} V(\theta) [\sigma^x(1+\cos\theta)+\sigma^y\sin\theta]\nonumber\\
&\times(i\sigma^y)[\sigma^x(1+\cos\theta)+\sigma^y\sin\theta]^T,\\
I^p=&-\kappa_c\Delta^p\int \frac{d\theta}{2\pi} V(\theta) [\sigma^x(1+\cos\theta)+\sigma^y\sin\theta]\nonumber\\
&\times(-\cos\theta\sigma^z+i\sin\theta)[\sigma^x(1+\cos\theta)+\sigma^y\sin\theta]^T.
\label{intp}
\end{align}
After some 
Pauli matrix algebra, we {can write}
\begin{align}
I_s=&\kappa_c\Delta^s~(i\sigma^y)\int \frac{d\theta}{2\pi} V(\theta)(2+2\cos\theta)\equiv V_0\kappa_c\Delta^s(\hat\k'),\nonumber\\
I_p=&\kappa_c\Delta^p~(-\sigma^z)\int \frac{d\theta}{2\pi} V(\theta)(2+2\cos\theta)\equiv V_0\kappa_c\Delta^p(\hat\k'),
\end{align}
where in the last step of each equation we used Eq.\ (\ref{sp2}) at $\theta=0,$ and we have defined
\begin{align}
V_0\equiv\int\frac{d\theta}{2\pi}V(\theta)(2+2\cos\theta).
\label{eqB}
\end{align}
 From this we see that the $s$-wave  and the $p$-wave channels are indeed degenerate. Combined with Eq.\ (\ref{eqA}), the  critical temperature is given by
\begin{align}
T_c=\Lambda e^{-{1}/({\kappa_cV_0})}=\Lambda\exp\(-\frac{4}{\lambda^2N(0)V_0}\)
\label{tc}
\end{align}
 for \emph{both} orders.
 
 Note that, the momentum dependence of the bosonic fluctuations $V(\theta)$ turns out to play no role in distinguishing the critical temperatures for the  $s$-wave and $p$-wave pairing channels. Thus, even though a degeneracy between these two channels was obtained in Ref.\ \onlinecite{Fu-2015} by approximating $V(\theta)$ as a constant, we have now shown that this conclusion applies for \emph{any} form of $V(\theta).$ This is surprising since, in many well-known cases of unconventional superconductivity,  
 the momentum dependence of the bosonic fluctuations typically has a strong effect on selecting the pairing symmetry. For example, in the $^3$He-A phase~\cite{Leggett-1975,Vollhardt-1990} the $p$-wave pairing channel is enhanced by ferromagnetic fluctuations peaking around ${\bm Q}=0$; in high-$T_c$ cuprate~\cite{Metlitski-2010} and Fe-pnictide~\cite{Chubukov-2012} superconductors the $d$-wave and $s_\pm$-wave pairing channels are enhanced by antiferromagnetic fluctuations peaking around ${\bm Q}=(\pi,\pi)$ and ${\bm Q}=(\pi,0)/(0,\pi)$, respectively. Also, in the candidate {chiral}  superconductor Sr$_2$RuO$_4$, it has been speculated that the $p$-wave pairing channel is enhanced by spin fluctuations with momentum ${\bm Q}=2k_F$ due to the nearly-nested $\beta$-band.\cite{Rice-1996,Raghu-2010} However, the situation {that} we have here is distinct from all of the examples above since we find a strong tendency for $p$-wave (or $s$-wave) no matter what the momentum structure of the bosonic modes is. We will show in the next Section that there is a deeper reason for this robust degeneracy, and uncover why the $p$-wave instability discussed here is not fine-tuned and does not rely on any peak structure of the bosonic susceptibility  in momentum space.
 
We can easily see that the degeneracy between the $s$- and $p$-wave SC orders extends to the case of a 3D spherical FS as well. In the gap equation Eq.\ \eqref{gap}, one can always define $\theta$ in a rotated $\hat {\bm x},\hat {\bm y}$ basis within the 2D plane formed by $\k$ and $\k'$, and the only difference in the 3D case would be that one needs to integrate over an additional  $\varphi$ angle, but rotational invariance implies that the resulting integrals are independent of $\varphi$ and the degeneracy persists.

Finally, we note that 
our analysis can be generalized to any odd-parity pairing channel of higher angular momenta $l=2n+1$. To obtain odd parity pairing instabilities with higher $l$'s, the corresponding QCP required is characterized by a $\bm d_\k$ with the same winding number $l$ over the FS. The analysis of the SC orders is similar to that above, and the only difference would be replacing $\cos\theta$ and $\sin\theta$ in Eqs.\ \eqref{sp2}-\eqref{eqB} with $\cos l\theta$ and $\sin l\theta$, and the $p$-wave order with a more general form
\begin{align}
{\mathcal H}_{\rm odd}=\Delta^{\rm odd} \; \hat{\bm d}_\k\cdot (i\bm\sigma \sigma^y)_{\alpha\beta}\; c^\dagger_{\k\alpha}c^\dagger_{-\k\beta}.
\end{align}
Therefore, in the vicinity of such a QCP, the $s$-wave channel  and odd-parity pairing channel with $l=2n+1$ are degenerate.
Without loss of generality, we simply use the case where $l=1$ (namely, the $p$-wave case) as an example of the odd-parity pairing for the rest of this article, and our conclusions naturally apply for QCP's with any $l$.

So far, we have considered a system close to but  away from the QCP. The full quantum-critical pairing problem is more complicated~\cite{Millis-1988,Scalapino-1995,acf-2001,Metlitski-2010,Efetov-2013,Wang-2013,Metlitski-2015,Schattner-2015,Raghu-2015} in the sense that (i) bosonic and fermionic self-energies and vertex corrections can have non-analytic behavior and generally cannot be neglected, (ii) the Eliashberg approximation which confines the important fermionic degrees of freedom to the vicinity of the FS is generally invalid, and (iii) the contribution from non-ladder diagrams is generically comparable to that from the ladder diagrams. These issues have been studied and addressed in previous works~\cite{acf-2001,Abanov-2003,Metlitski-2010,Wang-2013,Wang-2013a,Wang-2015,Wang-2015a,Metlitski-2015,Raghu-2015} for different quantum critical pairing problems within a large-$N_f$ framework, where $N_f$ corresponds to the number of fermionic flavors. In particular,  it was found that the Eliashberg approximation and the ladder approximation become exact in the $N_f\to \infty$ limit. Within our work, we refrain from presenting a full quantum-critical analysis, but instead simply assume the degeneracy described above holds in the critical regime, at least approximately.

\section{The $\U\times\U$ symmetry}

\subsection{Free energy of the superconducting order parameters}

{We begin by considering first} the simplest case of a $p$-wave inversion-breaking order that yields a degeneracy between $s$-wave and $p$-wave superconducting order as an example. Due to this degeneracy, near the critical temperature for the superconducting thermodynamic phase transition the free energy has the form $F=\alpha(|\Delta_s|^2+|\Delta_p|^2)+O(\Delta^4)$ at quadratic order.

\begin{figure}[t]
\includegraphics[width=0.45\textwidth]{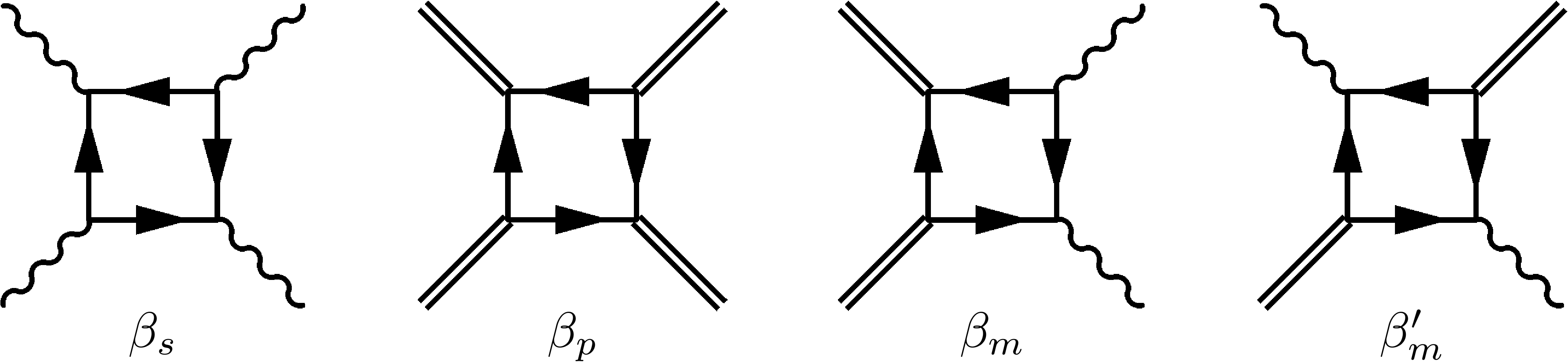}
\caption{The diagrams for $\beta$ coefficients in which $\Delta_s$ is represented by wavy lines and $\Delta_p$ by double lines. The fermion lines merging at a given vertex carry opposite frequencies and momenta.}
\label{beta}
\end{figure}

At quartic order, the free energy is generically given by
\begin{align}
F=&\alpha(|\Delta_s|^2+|\Delta_p|^2)+\beta_s|\Delta_s|^4+\beta_p|\Delta_p|^4\nonumber\\
&+\beta_m|\Delta_s|^2|\Delta_p|^2+\beta_m'[\Delta_s^2(\Delta_p^*)^2+\Delta_p^2(\Delta_s^*)^2].
\label{15}
\end{align}
For generic coefficients,  this free energy has the expected global $\U$ symmetry of a superconductor.
For our case, the   coefficients $\beta_s$, $\beta_p$, $\beta_m$ and $\beta_m'$ in Eq.\ \eqref{15} can be directly calculated and are given by the square {Feynman} diagrams shown in Fig.\ \ref{beta}. {Their explicit expressions are} 
\begin{align}
&\beta_s=\frac{\beta}{4}\Tr[(i\sigma^y)(i\sigma^y)^\dagger(i\sigma^y)(i\sigma^y)^\dagger]=\frac{\beta}{2} \nonumber\\
&\beta_p=\frac{\beta}{4}\Tr[(i\hat\k\cdot{\bm \sigma}\sigma^y)(i\hat\k\cdot{\bm \sigma}\sigma^y)^\dagger(i\hat\k\cdot{\bm \sigma}\sigma^y)(i\hat\k\cdot{\bm \sigma}\sigma^y)^\dagger]=\frac{\beta}{2}\nonumber\\
&\beta_m=\beta\Tr[(i\hat\k\cdot{\bm \sigma}\sigma^y)(i\hat\k\cdot{\bm \sigma}\sigma^y)^\dagger(i\sigma^y)(i\sigma^y)^\dagger] =2\beta\nonumber\\
&\beta_m'=\frac{\beta}{4}\Tr[(i\hat\k\cdot{\bm \sigma}\sigma^y)(i\sigma^y)^\dagger(i\hat\k\cdot{\bm \sigma}\sigma^y)(i\sigma^y)^\dagger]=\frac{\beta}{2},
\end{align} 
where $\beta$ is the momentum and frequency integral over the four Green functions, given by
\begin{align}
\beta=\sum_{m,\k}G^2(\omega_m,\k)G^2(-\omega_m,-\k)=\sum_{m,\k}\frac{1}{(\omega_m^2+\epsilon_\k^2)^2},
\label{eq:beta}
\end{align}
and $\epsilon_{\k}$ is the fermionic dispersion. {Eq. \eqref{eq:beta} yields a temperature dependence of $\beta$, both for a 2D or 3D FS, that goes as $1/T^2$}, however the exact numerical coefficient  is not of particular interest to us here.

The free energy in terms of $\Delta_s$ and $\Delta_p$ up to quartic order is then
\begin{align}
F=&\alpha(|\Delta_s|^2+|\Delta_p|^2)+\frac\beta2(|\Delta_s|^4+|\Delta_p|^4)\nonumber\\
&+2\beta|\Delta_s|^2|\Delta_p|^2+\frac{\beta}2[\Delta_s^2(\Delta_p^*)^2+\Delta_p^2(\Delta_s^*)^2].
\label{eq:1}
\end{align}
This free energy has an additional  symmetry that can be made more transparent if we define
\begin{align}
\Delta_1=&(\Delta_s+\Delta_p)/\sqrt2 \nonumber\\
\Delta_2=&(\Delta_s-\Delta_p)/\sqrt2.
\label{19}
\end{align}
After some algebra, one can rewrite Eq.\ \eqref{eq:1} in  {the much} simpler form
\begin{align} 
F=&\alpha(|\Delta_1|^2+|\Delta_2|^2)+\beta(|\Delta_1|^4+|\Delta_2|^4)+\ldots
\label{20}
\end{align}
Hence, we see that, at least to quartic order, $\Delta_1$ and $\Delta_2$ decouple, and the symmetry of the free energy is  {actually} $\U\times \U$. Next we will analyze the origin of this decoupling, and show that it actually holds to all orders in the Ginzburg-Landau expansion.

\subsection{Origin of the enlarged symmetry}

The degeneracy of $s$-wave and $p$-wave pairing, and the enlarged $\U\times\U$ symmetry is not accidental, and holds beyond {a low order} perturbative expansion in $\Delta$'s. To see this, observe that  {the Hamiltonian $\mathcal{H}_{\rm SC}$, given by  Eq.\ \eqref{sp} and Eq.\ \eqref{19}, is}
\begin{align}
\mathcal{H}_{\rm SC}\equiv&\mathcal{H}_s+\mathcal{H}_p\nonumber\\
=&\sqrt2\Delta_1c^\dagger_\k \(\frac{1+\hat\k\cdot{\bm \sigma}}2\) (i\sigma^y)(c_{-\k}^{\dagger})^T\nonumber\\
&+\sqrt2\Delta_2c^\dagger_\k\( \frac{1-\hat\k\cdot{\bm \sigma}}2\) (i\sigma^y)(c_{-\k}^{\dagger})^T,
\end{align}
where $c_\k^\dagger=(c_{\k\uparrow}^\dagger,c_{\k\downarrow}^\dagger)$.
We define the fermionic helicity $\chi_{\k}\equiv {\hat \k}\cdot{\bm \sigma}$
and the helicity projection operator
\begin{align}
P_{\pm}(\k)\equiv\frac{1\pm\chi_{\k}}{2}.
\end{align}
{This  operator projects the single-particle  fermion states, created by $c^\dagger_{\k}$, onto helicity eigenstates} with $\chi_{\k}=\pm 1$, namely
\begin{align}
c_{\k,\pm}\equiv P_{\pm}(\k) c_{\k},~~ \chi_{\k} c_{\k,\pm}=\pm c_{\k,\pm}.
\end{align}
It is easy to show that $P_\pm(\k)$ satisfies $P_{\pm}(\k)^2=P_{\pm}(\k)$ and $P_\pm(\k)\sigma^y=\sigma^yP_{\pm}(\k)^T$.
Using these properties, we further obtain that
\begin{align}
\mathcal{H}_{\rm SC}=&\sqrt2\Delta_1c^\dagger_{\k,+} (i\sigma^y)(c_{-\k,+}^{\dagger})^T\nonumber\\
&+\sqrt2\Delta_2c^\dagger_{\k,-} (i\sigma^y)(c_{-\k,-}^{\dagger})^T,
\end{align}
where $c_{\k,\pm}\equiv P_{\pm}(\k) c_{\k}$.
Therefore, we identify $\Delta_1$ and $\Delta_2$ as the superconducting order parameters that couple to fermions with helicity $\pm 1$, respectively.

It turns out that the {{\em helical pairing fields}} $\Delta_1$ and $\Delta_2$ enable a more straightforward understanding of the enlarged symmetry, even though $\Delta_s$ and $\Delta_p$ are more physically transparent. 
We first notice that the boson-fermion vertex for fermions on the FS with momentum $\k$ and $\k+{\bm q}$, given by Eq.\ \eqref{eq4}, can be rewritten as
\begin{align}
\bar Q({\bm q})=&\frac12\sum_{\bm k} c^\dagger_{{\bm{k+q}} \alpha}[({\bm d}_{\bm k}+{\bm d}_{{\bm{k+q}}}) \cdot {\bm\sigma}_{\alpha\beta}]c_{\bm k\beta}
\nonumber\\
=&\frac\lambda2\sum_{\bm k} c^\dagger_{{\bm{k+q}} \alpha}(\chi_{\k}+\chi_{{\bm{k+q}}})_{\alpha\beta}c_{\bm k\beta},
\label{eq27}
\end{align}
where we have used ${\bm d}_\k=\lambda \hat \k$ on the FS. It is  clear that the vertex $\bar Q({\bm q})$ only couples fermions with the same helicity, since otherwise $\chi_\k+\chi_{\k+{\bm q}}=0$. Hence, the fermions in different helicity sectors completely {\it decouple}.

In the helicity basis, the interaction vertex has a much simpler form
\begin{align}
\bar Q({\bm q})
=&\lambda\sum_{\bm k\alpha} \[(c^\dagger_{{\bm{k+q}},+})_\alpha (c^{\phantom{\dagger}}_{\bm k,+})_\alpha-(c^\dagger_{{\bm{k+q}},-})_{\alpha}(c^{\phantom{\dagger}}_{\bm k,-})_{\alpha}\]\nonumber\\
=&\lambda\sum_{\k}\cos(\theta/2)\[c^\dagger_{\k+\bm q,+} c^{\phantom{\dagger}}_{\k,+}e^{i\varphi/2}+c^\dagger_{\k+\bm q,-} c^{\phantom{\dagger}}_{\k,-}e^{-i\varphi/2}\].
\end{align}
In the last line we used the fact that the spin orientations of, say, $c^\dagger_{\bm{k}+\bm{q},+}$ and $c_{\bm k,+}$ are different, hence the inner product in the (pseudo-)spin space is $\cos(\theta/2) e^{i\varphi/2}$, where $(\theta,\varphi)$ characterizes the angles between $\k$ and $\k+\bm q$ in 3D.

We can now obtain the linearized gap equations for the helical pairing fields $\Delta_1$ and $\Delta_2$. Within the Eliashberg approximation,  the important fermionic degrees of freedom  are close to the FS. After integrating over momenta normal to the FS and over frequency, we have
\begin{align}
\Delta_{1,2}=&-4\kappa_c\int\frac{d\theta}{2\pi}{V(\theta)}\cos^2(\theta/2)  \Delta_{1,2}.
\end{align}
Recalling Eq.\ \eqref{eqB}, we obtain a critical temperature $T_c$ that is identical to that in Eq.\ \eqref{tc}.
Indeed, since $\Delta_1$ and $\Delta_2$ are just linear combinations of $\Delta_s$ and $\Delta_p$, their critical temperatures have to be the same.
 
To summarize, in this subsection we have shown that: (i) the fermionic interaction mediated by the fluctuations of the inversion-breaking order in Eq.\ (\ref{V}) is diagonal in the helicity basis, and (ii) fermions within each helicity sector form Cooper pairs (leading to pairing fields $\Delta_1$ and $\Delta_2$) separately. Therefore, each pairing field has an independent $\U$ symmetry. This is the origin of the $\U\times\U$ symmetry in Eq.\ (\ref{20}). From the argument above we also see that the decoupling of $\Delta_1$ and $\Delta_2$ will hold at all orders of the Ginzburg-Landau expansion.
 
\section{Lifting (and restoring) the enlarged symmetry}

The decoupling between $\Delta_1$ and $\Delta_2$ relies on the fact that the boson-fermion interaction vertex induced by the inversion-breaking QCP only couples fermions with the same helicity. We {now} want to consider the stability of this symmetry in a realistic system. We expect that this enlarged symmetry generically will only be approximate, and, {except for some fine-tuned cases,} the $\U\times\U$ symmetry is {explicitly} broken down to a single $\U$ symmetry.

We can argue for this breakdown as follows. To begin with, we have not yet addressed the effect of the screened Coulomb interaction. The Coulomb interaction acts in the density-density channel, and thus  generically couples fermions with opposite helicities, inducing an effective Josephson tunneling term (between the two species of fermions) that will take the form $\sim\Delta_1\Delta_2^*$ in the free energy. This term breaks the $\U\times\U$ symmetry by locking the relative phase of the two components of the superconducting order parameter. The repulsive nature of the screened Coulomb interaction prefers the system to have a sign change in the gap function, since this will cancel the onsite wave function of the Cooper pair and avoids the short-range repulsion. Thus, up to quartic order {terms}, the free energy becomes
\begin{align}
F=&\alpha(|\Delta_1|^2+|\Delta_2|^2)\nonumber\\
&+\alpha_C(\Delta_1\Delta_2^*+\Delta_1^*\Delta_2)+\beta(|\Delta_1|^4+|\Delta_2|^4),
\end{align} where $\alpha_C>0.$ A simple analysis of this free energy yields that for the ground state, $\Delta_1=-\Delta_2$, and from Eq.\ (\ref{19}), we see that this represents a pure $p$-wave state.

Another important complication arises in the ordered phase of the inversion-symmetry-breaking order parameter $\phi$. In this case the FS splits into two with Fermi momenta $k_{F1}$ and $k_{F2}$ respectively, and each having a fixed helicity $\chi_{\k}=\pm1$ on the entire FS. To leading order, $k_{F1}-k_{F2}\propto\langle\phi\rangle$. The helical pairing fields $\Delta_1$ and $\Delta_2$ are still well-defined with the two split FS's, and in the ordered state $\langle \phi\rangle\neq 0$, the fluctuations of the $\phi$ field about its mean-field value mediate an attractive interaction between $\Delta_{1}$ and $\Delta_{2}$. However, in this case the cancellation of inter-helicity coupling in the boson-fermion vertex, namely Eq.\ (\ref{eq27}), does not hold. 

Let us take a closer look at this {question}. For the case of $l=1$, generically ${\bm d_\k}=\lambda(|\k|)\hat\k$, but the values of $\lambda$ for the low-energy fermions on the two FS's are not the same. Denoting $\lambda(k_{F1})=\lambda_1$ and  $\lambda(k_{F2})=\lambda_2$, we have $\lambda_1-\lambda_2\propto k_{F1}-k_{F2}\propto \langle\phi\rangle$, and for the fermionic part of the vertex,
\begin{align}
\bar Q({\bm q})
&=\sum_{\bm k\alpha} \left\{\lambda_1(c^\dagger_{{\bm{k+q}},+})_\alpha (c_{\bm k,+})_\alpha-\lambda_2(c^\dagger_{{\bm{k+q}},-})_{\alpha}(c_{\bm k,-})_{\alpha}\right.\nonumber\\
& \left.+\frac{\lambda_1-\lambda_2}{2}[(c^\dagger_{{\bm{k+q}},+})_\alpha (c_{\bm k,-})_\alpha-(c^\dagger_{{\bm{k+q}},+})_\alpha (c_{\bm k,-})_\alpha]\right\}, \nonumber\\
\end{align}
which is no longer diagonal in the helical basis. The result of the residual {off-diagonal} interaction, which now can couple fermions with opposite helicities, is that $\Delta_1$ and $\Delta_2$  are again linearly coupled. However, in this case this residual interaction is \emph{attractive} in the pairing channel, just like the 
pairing interactions
 within each helical sector. 
 Thus it prefers $\Delta_1$ and $\Delta_2$ to be of the same sign, which will maximize the condensation energy. 

Now, together with the effect of the Coulomb interaction, the free energy for $\Delta_{1}$ and $\Delta_2$ in the ordered phase, $\langle \phi \rangle \neq 0$,  becomes
\begin{align}
F=&\alpha_1|\Delta_1|^2+\alpha_2|\Delta_2|^2+(\alpha_C-\alpha_S)(\Delta_1\Delta_2^*+\Delta_1^*\Delta_2)\nonumber\\
&+\beta_1|\Delta_1|^4+\beta_2|\Delta_2|^4+\ldots,~~~~~~~\alpha_{C,S}>0.
\label{eq32}
\end{align}
The magnitude of $\alpha_S$ is larger for a larger inter-helicity interaction strength, {which is proportional to $(\lambda_1-\lambda_2)^2$, and hence} increases as the FS splitting increases, i.e., as $\langle\phi\rangle$ increases.
The coefficients $\alpha_{1}$, $\alpha_2$, $\beta_{1}$ and $\beta_2$ can be obtained by a Hubbard-Stratonovich transformation and then integrating out the low-energy fermions with both helicities.
Using the standard technique, we obtain
\begin{align}
\alpha_{1}=\frac{1}{g_{1}}-\sum_{m,\k}\frac{1}{\omega_m^2+\epsilon_{1\k}^2},~~&
\alpha_{2}=\frac{1}{g_{2}}-\sum_{m,\k}\frac{1}{\omega_m^2+\epsilon_{2\k}^2},\nonumber\\
\beta_{1}=\sum_{m,\k}\frac{1}{(\omega_m^2+\epsilon_{1\k}^2)^2},~~&\beta_{2}=\sum_{m,\k}\frac{1}{(\omega_m^2+\epsilon_{2\k}^2)^2},
\label{eq33}
\end{align}
where $g_{1,2}\propto\lambda_{1,2}$ are the effective couplings in the SC channel for the FS with helicity $\pm1$ respectively, and the two fermionic dispersions can be approximated by $\epsilon_{1\k}\simeq {v}_{F}(|\k|-k_{F1})$ and $\epsilon_{2\k}\simeq {v}_{F} (|\k|-k_{F2})$ respectively. 

Note that the same free energy as Eq.\ \eqref{eq32} can also be derived in terms of $\Delta_s$ and $\Delta_p$. However the derivation is much more tedious since: (i) the Green functions in a general basis have a matrix form, and (ii)  the inversion symmetry is broken, and hence many more terms, such as $\Delta_s\Delta_p^*$ and $|\Delta_s|^2\Delta_s\Delta_p^*$, are allowed in the free energy. As a check we have verified that it leads to the same result as in Eq.\ \eqref{eq32}, and present the technical details in Appendix \ref{app:a}.

The ground state configuration of Eq.\ \eqref{eq32} depends now on the sign of $\alpha_C-\alpha_S$. For $\alpha_C>\alpha_S$, the relative phase between $\Delta_1$ and $\Delta_2$ is $\pi$. Note that since $\alpha_1\neq\alpha_2$ and $\beta_1\neq\beta_2$, the magnitudes of $\Delta_1$ and $\Delta_2$ are different. Simple {algebra} shows that, in terms of the $s$-wave and $p$-wave order parameters, this corresponds to a state where $\Delta_s=\varepsilon\Delta_p$, where {$\varepsilon$ is a {\it real} number (hence time-reversal is not broken)} and $|\varepsilon|<1$. We denote this state as a $p+\varepsilon s$ state. This mixing of $s$- and $p$-wave channels~\cite{Agterberg-2004} can be viewed as a result of the broken inversion symmetry due to $\langle\phi\rangle\neq0$. On the other hand, for $\alpha_C<\alpha_S$, the phase difference between $\Delta_1$ and $\Delta_2$ in the ground state is 0, and this corresponds to $\Delta_p=\varepsilon\Delta_s$. We denote this state as $s+\varepsilon p$. We show in the next Section that the $p+\varepsilon s$ state is a topological superconducting state, while the $s+\varepsilon p$ state is trivial.

Interestingly, when $\alpha_C=\alpha_S$, i.e., at the phase boundary between $p+\varepsilon s$ and $s+\varepsilon p$, in Eq.\ \eqref{eq32}, $\Delta_1$ and $\Delta_2$ decouple from each other,
and the $\U\times\U$ global symmetry is restored in the free energy. Since $\alpha_1\neq \alpha_2$, there exist two mean-field transition temperatures $T_1$ and $T_2$, corresponding respectively to the onset of $|\Delta_1|\neq0$ and $|\Delta_2|\neq 0$. Note that, away from the phase boundary, only the \emph{higher} transition temperature 
{matters since}, away from the $\alpha_C=\alpha_S$ line, $\Delta_1$ and $\Delta_2$ are coupled, and either one of them developing a nonzero magnitude immediately induces the other one.  Without loss of generality, we assume that $T_1>T_2$.

\begin{figure}[hbt]
\includegraphics[width=0.8\columnwidth]{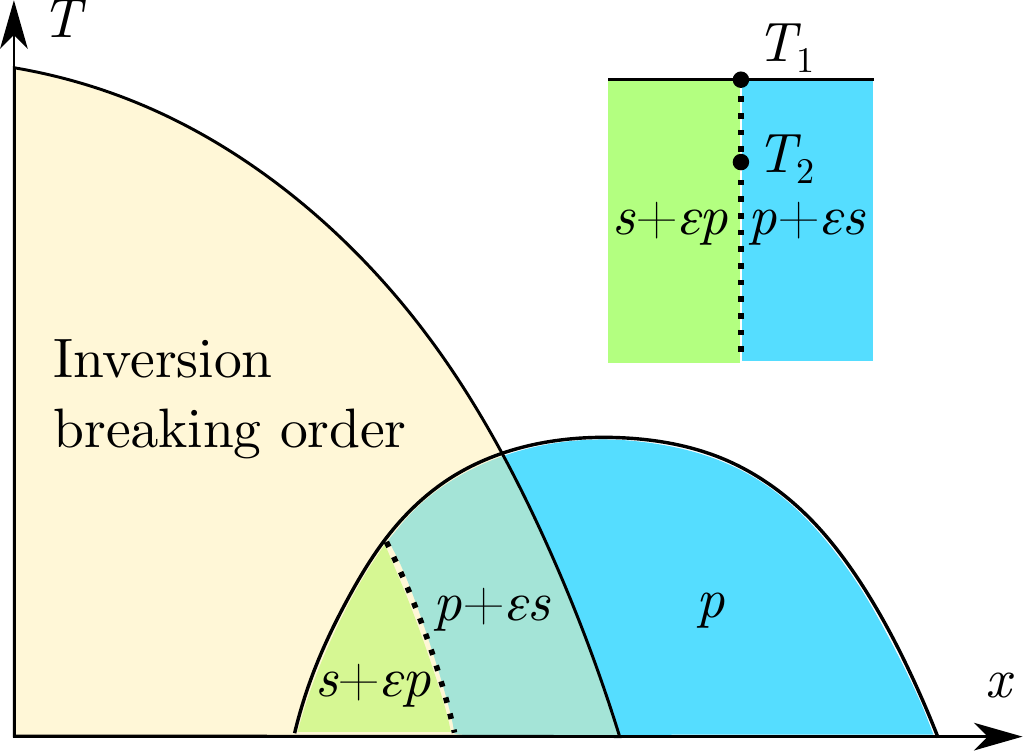}
\caption{The mean-field phase diagram for the inversion symmetry breaking order and the superconducting orders. Main figure: The $x$ axis is an arbitrary parameter controlling the onset of the inversion symmetry breaking order. The pairing symmetry of the superconducting orders are labeled. The dotted line separate phases with the same symmetry but with different topological classification. Close to this dotted line the system has an approximate $\U\times\U$ symmetry. For simplicity we have presumed the phases extend to the quantum-critical regime. Inset: Details of the phase diagram at the topological phase boundary. Due to the $\U\times\U$ symmetry there exist two transition temperatures, $T_1$ and $T_2$.}
\label{phase}
\end{figure}

We can summarize the analysis in this Section in {the} mean-field phase diagram shown schematically in Fig.\ \ref{phase}. On the disordered side of the inversion symmetry breaking order $\phi$, the SC order is of $p$-wave symmetry, and is hence a time-reversal invariant topological superconductor. On the ordered side, the SC order is an admixture of $p$-wave and $s$-wave orders, and which one is larger in magnitude depends on the interplay between $\alpha_S$ and $\alpha_C$. As discussed above, $\alpha_S$ is an increasing function of the expectation value $\phi$, hence we expect an $p+\varepsilon s$ state immediately into the inversion breaking order phase, and a $s+\varepsilon p$ state deep into the ordered region (assuming that  the screened Coulomb interaction is approximately a constant strength). The $s+\varepsilon p$ state and $p+\varepsilon s$ state have an identical classification as far as symmetry is concerned, but as we will show below, they have distinct topological classifications. It is exactly at the phase boundary between $s+\varepsilon p$ and $p+\varepsilon s$ states that the $\U\times\U$ symmetry is restored in the free energy, and at which there generically exist two transition temperatures $T_{1,2}$ corresponding to the two $\U$'s.

 \begin{figure}[hbt]
 \includegraphics[width=0.5\columnwidth]{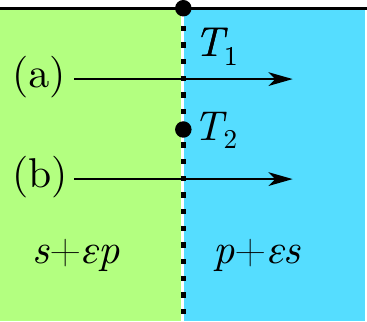}
 \caption{The phase transition between the trivial $s+\varepsilon p$ state and topological $p+\varepsilon s$ state. For $T_2<T<T_1$ [case (a)], the transition occurs  
 {with a vanishing SC order parameter $\Delta_2=0$, with $\Delta_1\neq 0$}, while for $T<T_2$ [case (b)] the transition occurs {
 with $\Delta_1\neq 0$ and $\Delta_2\neq0$. At $T=0$ there  is a quantum phase transition between these two superconducting states} without a gap closing of the fermionic quasiparticles but by going through a path that breaks effectively the time-reversal symmetry.}
 \label{phaseb}
 \end{figure}
 
\section{Properties of the superconducting state}

\subsection{Topological characterization of the superconducting states}

First, we discuss the topological properties of the superconducting states.  Both the  $p+\varepsilon s$ state and the $s+\varepsilon p$ state preserve time-reversal symmetry and have fully-gapped {fermionic} quasi-particles on their helical FS's. Thus, these states are candidates for  time-reversal invariant topological superconductors. For the 3D case, in the weak-coupling limit we are considering, their topology can be determined by computing the FS topological index\cite{Qi-2010a}
\begin{align}
N_{W} =  \frac{1}{2} \sum_{i=1,2} \sgn (\Delta_i) C_i , 
\label{nw}
\end{align}
where sgn($\Delta_{i}$) is the sign of the pairing field on the Fermi surface $i,$ and $C_i$ is the winding/Chern number of the Berry curvature piercing the closed Fermi surface. For the 2D case, by a dimensional reduction procedure, the topology is characterized by a {$\mathbb{Z}_2$ topological} invariant $N_{2D}=(-1)^{N_W}$.\cite{Qi-2010a}

 In our case, where ${\bm d}_\k \propto \k$ and the inversion breaking order has $p$-wave character $C_1=-C_2 = +1$ (without loss of generality). For the pure superconducting $p$-wave state and the $p+\varepsilon s$-wave state, the $\Delta_1$ and $\Delta_2$ fields on the two helical FS's have opposite signs, and hence $N_W=\pm 1.$ For the $s+\varepsilon p$ state, $\Delta_{1,2}$ have the same sign and $N_W=0$. Therefore, in both 2D and 3D cases, the system is a time-reversal invariant topological superconductor in the
 $p$ and $p+\varepsilon s$ states, while it is topologically trivial in the $s+\varepsilon p$ state. {We note in passing that, strictly speaking, the topological indices are sharply defined only at $T=0$, but we presume that the topological properties of the superconducting states at $T=0$ survive at finite temperature.}

The phase boundary between the $s+\epsilon p$ and $p+\epsilon s$ states characterizes a phase transition where the topological index changes. For $T_2<T<T_1$, we have $\Delta_1\neq 0$ and $\Delta_2 =0$ on the phase boundary, which means that at the ``topological" phase transition, the SC {order parameter} on one of the two FS's vanishes. However, interestingly, for $T<T_2$, both $\vert\Delta_{1,2}\vert\neq0$ at the phase boundary and {the transition between the topological and trivial states occurs without a vanishing SC order parameter. Extrapolating to $T=0$, we see that at this phase boundary the system undergoes a topological phase transition without a gap closing of the fermionic quasi-particles.\cite{Nagaosa-2013}} This is possible because the relative phase of $\Delta_1$ and $\Delta_2$ is not fixed on this boundary, and there is a gapless bosonic critical mode present. {This critical mode can be regarded as a  Leggett mode,\cite{Leggett-1966} which plays the role of a Goldstone mode of the additional spontaneously broken $U(1)$ phase symmetry at the phase boundary} (see Sec.\ \ref{leggett} for details). In this regime of temperature, one can have a path connecting the topological state to the trivial state where the relative phase can smoothly rotate from 0 to $\pi$. However, time reversal symmetry would be broken along this path, and the topological index in Eq.\ \eqref{nw} is ill-defined.\cite{Nagaosa-2013} {Thus, in this second mechanism,  the transition between a topological superconductor and a trivial one circumvents the vanishing of the SC order parameter.} We illustrate the two topological phase transition paths with and  without fermionic gap closing in Fig.\ \ref{phaseb}.

\subsection{Topological defects at the topological phase boundary}
Along the topological phase boundary, the system has a $\U\times\U$ global symmetry in the free energy. Owing to this enlarged symmetry, various topological defects can now exist. 

For the 2D case, the system supports two types of superconducting vortices. The first type is a fractional vortex, where only one of the the SC order parameters has a phase winding.\cite{Volovik_1999,ivanov} In this case the magnetic flux through the vortex is a fractional multiple of $h/2e.$ To see this, we write down the free energy in terms of the two phase modes
\begin{align}
F=\frac{\rho_1}{2}\left({\bm \nabla}\phi_1-2e{\bm A}\right)^2+\frac{\rho_2}{2}\left({\bm \nabla}\phi_2-2e{\bm A}\right)^2,
\label{35a}
\end{align}
where $\rho_{1,2}\propto|\Delta_{1,2}|^2$ are the superfluid stiffnesses of the two superconducting components, and $\Delta_{1,2}=|\Delta_{1,2}|\exp(i\phi_{1,2})$. For simplicity, first consider the case $\rho_1=\rho_2\equiv \rho$, where the free energy takes the even simpler form
\begin{align}
F=\frac\rho4[{\bm \nabla}(\phi_1+\phi_2)-4e{\bm A}]^2+\frac\rho4[{\bm \nabla}(\phi_1-\phi_2)]^2.
\label{35}
\end{align}
For a vortex in which only one phase angle winds, say $\phi_1$, we can easily see that the magnetic flux through this vortex is $\Phi=hc/4e$,  one half of the standard superconducting flux quantum $\Phi_0=hc/2e$. In this case, this is just a half-quantum vortex. 

For our case, the $\U\times\U$ symmetry occurs in the inversion broken phase, and $\rho_1\neq\rho_2$. Then we can rewrite Eq.\ \eqref{35a} as
\begin{align}
F=&\frac{\rho_1+\rho_2}{2}\[\frac{\rho_1}{\rho_1+\rho_2}{\bm \nabla}\phi_1+\frac{\rho_2}{\rho_1+\rho_2}{\bm \nabla}\phi_2-2e{\bm A}\]^2\nonumber\\
&+\frac{\rho_1\rho_2}{2(\rho_1+\rho_2)}\[{\bm \nabla}(\phi_1-\phi_2)\]^2
\label{35b}
\end{align}
 A simple calculation\cite{babaev} shows that for vortices in $\phi_1$ and $\phi_2$ alone, the fluxes are respectively
\begin{align}
\Phi_1=\frac{\Phi_0\rho_1}{\rho_1+\rho_2},~~
\Phi_2=\frac{\Phi_0\rho_2}{\rho_1+\rho_2},
\end{align} 
i.e., both $\Phi_1$ and $\Phi_2$ are fractional, and their sum is the full flux quantum. From the second term of Eq.\ \eqref{35b} we see the fractional vortices are logarithmically confined. Away from the phase boundary between $p+\varepsilon s$ and $s+\varepsilon p$, these fractional vortices become linearly confined, but they are expected to still exist if one presumes that the Coulomb interaction and FS splitting are weak. This analysis is similar to what is found in pair-density-wave phases\cite{Berg-2009,Berg-2009b,Agterberg-2008} and in $p_x+ip_y$ superconductors.\cite{Vakaryuk-2009} 

A second type of vortices are the conventional  full quantum vortices, described by a simultaneous winding of both $\phi_1$ and $\phi_2$ around a defect. The penetrating flux is the conventional value of $\Phi_0=hc/2e$. The vortex-vortex interaction, as can be seen from Eq.\ \eqref{35b}, is screened by $\bm A$ at length scales larger than the penetration length $\lambda$. 

In two spatial dimensions, there can be no continuous symmetry breaking {at finite temperatures} due to phase fluctuations. However, below the two mean-field temperatures, the phases of the superconducting order parameters can develop quasi-long-range order via Berezinskii-Kosterlitz-Thouless (BKT) transitions. In the type-II limit where the screening length $\lambda$ is much smaller than the size of the vortex core $\xi$, the screening effect can be neglected, and each one of $\phi_1$ and $\phi_2$ goes through a BKT transition. In this case the low temperature phase is a superconducting one. However, it was found in Ref.\ \onlinecite{babaev} that in the (opposite) type-I limit where $\lambda\ll\xi$, the first term of \eqref{35b} is completely screened. The second term in Eq.\ \eqref{35b} then gives a {\it single} BKT transition temperature
\begin{align} 
T_{BKT}=\frac{\pi\rho_1\rho_2}{4(\rho_1+\rho_2)},
\end{align}
where we have assumed $\rho_{1,2}$ are constants in temperature. Below this temperature the relative phase $\phi_1-\phi_2$ is quasi-long-range ordered, while $\phi_{1,2}$ individually remain disordered. Thus, the system exhibits a non-superconducting quasi-superfluid behavior.\cite{babaev} In this case the fractional vortices combine into full ones, but the latter remain deconfined. 

It is well-known that the core of  vortices can trap localized, fermionic bound states. For a fractional vortex, at its core there is a phase winding of only one of the pairing fields, and the other pairing field remains smooth. Because of the non-trivial texture of the Fermi surface, one can easily deduce that the half quantum vortex will localize a single Majorana zero mode,\cite{Fu-2008,Roy-2010} which has non-Abelian braiding statistics. The core of the full quantum vortex traps two zero modes. Each zero mode originates from one of the Fermi surfaces with winding number $\pm1$. In the absence of time-reversal symmetry, the two zero modes can interact each other and get split away from the zero energy. Thus the full vortex has  trivial statistics. 

In three spatial dimensions, the $\U\times\U$ symmetry also supports exotic topological defects.  It was found in Ref.\ \onlinecite{babaev2} that the $\U\times\U$ model maps onto an O(3) nonlinear $\sigma$-model, and hence supports knot solitons.\cite{faddeev} It would be interesting to see in the current context if the core of the knot soliton traps exotic zero modes. We leave this to  future work.

\subsection{Unconventional Josephson effects}

Owing to its unconventional pairing symmetry and  nontrivial topology, our system will exhibit unconventional Josephson effects when tunneling to a conventional $s$-wave superconductor. First, on the disordered side of the inversion breaking order, the system is a $p$-wave time-reversal invariant topological superconductor, which has localized Majorana modes at the junction. It has been found in Ref.\ \onlinecite{Chung-2013} that in two and three spatial dimensions the tunneling to the s-wave superconductor due to the Majorana modes generates a Josephson current that has a periodicity of $\Delta\theta=\pi$, half of the conventional one. On the other hand, the ``regular" Josephson coupling, i.e., the Josephson effect not mediated by Majorana boundary modes,  can only occur through a two-pair hopping process, due to the odd parity of the $p$-wave superconductor. As a result, the periodicity of this regular Josephson current also has a periodicity $\Delta\theta=\pi$, although its amplitude would be suppressed because the tunneling process is of higher order. 
The latter is similar to the recent experimental prediction on Josephson tunneling between a conventional superconductor and a pair-density-wave superconductor, which breaks translational symmetry.\cite{Berg-2009}

 On the ordered side of the inversion breaking order, however, the pairing symmetry is $p+\varepsilon s$ or $s+\varepsilon p$, which always has an $s$-wave component. Hence, we expect the dominant tunneling current will always have a regular period of $\Delta\theta=2\pi.$ This is in addition to the possible period-$\pi$ one that arises from the tunneling of Majorana bound states if the SC state is topological.
 
 \subsection{Leggett mode close to the topological phase boundary}
 \label{leggett}
The two $\U$ fields $\Delta_1$ and $\Delta_2$ are only coupled by a quadratic term $(\alpha_C-\alpha_S)$, which is small in the magnitude of the screened Coulomb interaction strength and the expectation value of the inversion breaking field $\phi$. This coupling term locks the relative phase $\phi_{1}-\phi_2$ between $\Delta_1$ and $\Delta_2$ in the ground state at 0 or $\pi$. But as long as this coupling is sufficiently small, namely $\alpha_{C,S}\ll \alpha_{1,2}$, there is a low energy Leggett mode~\cite{Leggett-1966} lying inside the superconducting gap and corresponding to the fluctuations of $\phi_{1}-\phi_2$ from its minimum value. The Leggett mode has been predicted to exist in superconductors with two SC gaps, e.g., MgB$_2$,\cite{Klein-2010} however it has not yet been observed, presumably due to the strong coupling between the two gaps. However, in our case it would be interesting to test the existence of a vanishing Leggett gap {at} the phase boundary between $s+\varepsilon p$ and $p+\varepsilon s$ phases, where such a coupling is small and  tuned to vanish. {In other words, as noted in the previous section, the Leggett mode becomes the Goldstone boson of the spontaneously broken additional $U(1)$ symmetry that exists at the phase boundary between the $p+\varepsilon s$ and $s+\varepsilon p$ superconducting phases when below the (lower) critical temperature $T_2$.}

\section{Conclusions}

In this work we have analyzed in detail the pairing instability in the vicinity of an inversion-symmetry breaking order as a mechanism for odd-parity superconductivity, which is helpful for the realization of a time-reversal invariant topological superconductor.  

We found that, as a result of an emergent $\U\times\U$ symmetry, there are two degenerate superconducting channels: a conventional $s$-wave channel, and  a time-reversal invariant odd-parity channel (in the simplest case, a $p$-wave channel). We showed that the enlarged $\U\times\U$ symmetry emerges from  fermions with opposite helicities that form Cooper pairs independently of each other. The $\U\times\U$ symmetry enables an exotic superconducting vortex, namely a fractional quantum vortex, which binds a single Majorana zero mode at its core. On the other hand, in a realistic system the $\U\times\U$ can be lowered to a single $\U$ due to coupling between the fermions with opposite helicities. In particular, we discussed two scenarios for an inter-helicity coupling:  the Coulomb interaction, and Fermi-surface splitting due to inversion-symmetry breaking. The former tends to favor the $p$-wave state and the latter favors the $s$-wave state. Depending on the interplay between these effects we obtained a phase diagram for the SC orders, in which the pairing symmetry can be $p$, $p+\varepsilon s$, or $s+\varepsilon p$. Making use of a simple index theorem, we have identified the $p$ and $p+\varepsilon s$ states as topological superconducting states. In addition, we have also discussed other possible experimental implications, such as a Josephson tunneling experiment, and the existence of Leggett mode corresponding to the fluctuation of the locking angle between the two $\U$ pairing fields. 

One issue beyond the scope of this paper, which we leave as future work, is the pairing problem in the quantum-critical regime. As mentioned above, in this regime the ``normal state" generally becomes a non-Fermi liquid, and the pairing problem becomes rather involved.\cite{Millis-1988,Scalapino-1995,acf-2001,Metlitski-2010,Efetov-2013,Wang-2013,Metlitski-2015,Schattner-2015,Raghu-2015} It would be of theoretical and practical interest to analyze the interplay between the $s$-wave and odd-parity pairing channels in the presence of critical parity fluctuations. Particularly, in the case where $\bm \sigma$ represents the actual electron spin, it was found that the system is close to a Lifshitz multi-critical point\cite{Wu-2007} by tuning the external spin-orbit coupling. The possible non-Fermi liquid behavior and low temperature instabilities around this multi-critical point have not been considered before.
 Another interesting issue is to analyze the same QCP-mediated pairing problem when the continuous rotational symmetry is reduced to point group symmetries.\cite{Fu-2015b} In this case the $\U\times\U$ symmetry would be broken, but the interplay between the superconducting orders remains to be seen.

\begin{acknowledgements}
We thank Daniel Agterberg, Andrey Chubukov, Liang Fu, Tony Leggett and Peng Ye for useful discussions. This work was supported  in part by the Gordon and Betty
Moore Foundation's EPiQS Initiative through Grant No.
GBMF4305 at the University of Illinois (YW) and by the National Science Foundation through grants CAREER No. DMR-
1351895 (TLH) and No. DMR 1408713 at the University of Illinois (EF), and the Brain Korea 21 PLUS Project of Korea Government (GYC).  GYC thanks ICMT for their hospitality. 
\end{acknowledgements}

\appendix
\section{Alternative derivation of the free energy inside the inversion symmetry breaking order}
\label{app:a}

In this Appendix, we present another way of obtaining the free energy of the superconducting orders inside the inversion symmetry broken phase, namely Eq.\ \eqref{eq32}. Instead of  keeping the $\U\times\U$ symmetry explicit by using the order parameters $\Delta_{1,2}$, we stick with original parameters $\Delta_{s,p}$, which couple to fermions via Eq.\ \eqref{sp}. We show that the result agrees with Eq.\ \eqref{eq32}, although the analysis is significantly more tedious. 

In the ordered state of $\phi$, the Green function is renormalized to (assuming $\bm d_{\k}=\lambda(|\k|) \hat\k$)
\begin{align}
\hat G(\omega_m,\k)=&[i\omega_m-\epsilon_\k-\lambda(|\k|)\phi(\hat\k\cdot{\bm \sigma})]^{-1},\nonumber\\
=&\frac{i\omega_m-\epsilon_\k+\lambda(|\k|)\phi\hat\k\cdot{\bm \sigma}}{[i\omega_m-\epsilon_\k]^2-\lambda(|\k|)^2\phi^2}.
\end{align}
To simplify notations we drop the explicity $|\k|$ dependence in $\lambda(|\k|)$ hereafter.

We first focus our analysis in 2D. In the 3D case, we need to  integrate over an additional spherical coordinate $\varphi$ in the momentum space. However, by rotational invariance this would not lead to a different result than the 2D case.
Denoting the 2D momentum space by the $xy$ plane, we have in explicit form, 
\begin{align}
\hat G(\omega_m,\k)=\frac{-1}{[i\omega_m-\epsilon_\k]^2-\lambda^2\phi^2}\(\begin{array}{cc}
i\omega_m-\epsilon_\k&\lambda\phi e^{-i\theta} \\
\lambda\phi e^{i\theta}& i\omega_m-\epsilon_{\k}
\end{array}\),
\end{align}
where, we remind, $\theta$ is the angle on the FS [see Fig.\ \ref{gap}(b)].
One can then use this form of Green function to compute coefficients of the free energy.

As we said, in the absence of inversion symmetry, more terms are allowed into the free energy. Most generically the free energy takes the form
\begin{align}
F=&\bar\alpha_s|\Delta_s|^2+\bar\alpha_p|\Delta_p|^2+\alpha_{sp}(\Delta_s\Delta_p^*+\Delta_p\Delta_s^*)\nonumber\\
&+\bar\beta_s|\Delta_s|^4+\bar\beta_p|\Delta_p|^4\nonumber\\
&+\bar\beta_m|\Delta_s|^2|\Delta_p|^2+\bar\beta_m'[\Delta_s^2(\Delta_p^*)^2+\Delta_p^2(\Delta_s^*)^2] \nonumber\\
&+(\beta_{sp}|\Delta_s|^2+\beta_{sp}'|\Delta_p|^2)(\Delta_s\Delta_p^*+\Delta_p\Delta_s^*).
\label{16}
\end{align}
The standard procedure to compute the coefficients of the free energy is to apply a Hubbard-Stratonovich transformation to decouple the four-fermion interaction into two parts. The first part is quadratic in the bosonic order parameter field, whose coefficient is given by the inverse of the coupling strength in the corresponding channel. The second part is a Yukawa-type term between the bosonic field and the fermionic bilinear term. The fermionic degrees of freedom in the second part can be integrated out, resulting in quadratic and higher order terms in the bosonic fields in the free energy.

\begin{figure}[h]
\includegraphics[width=0.8\columnwidth]{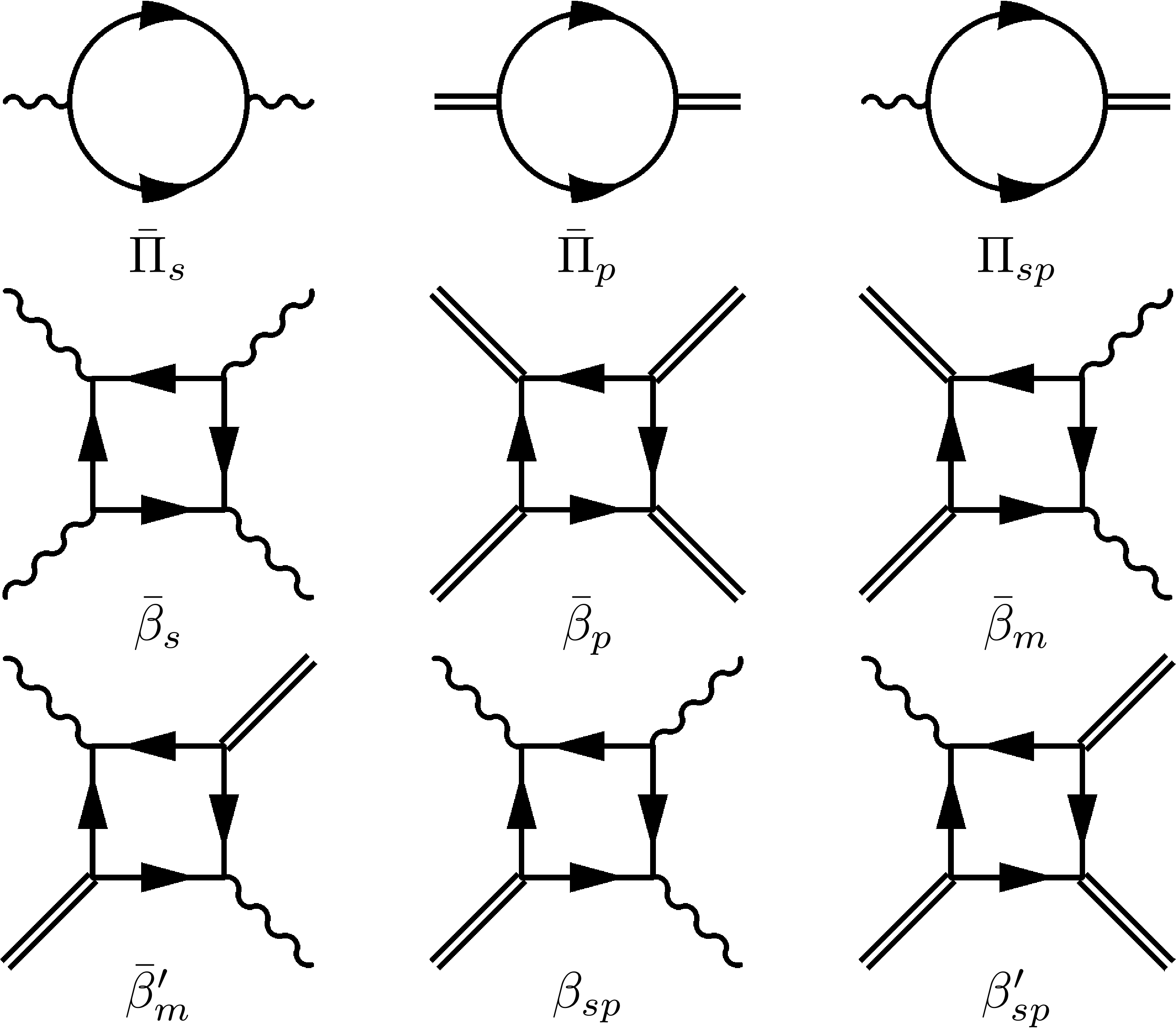}
\caption{The diagrams for coefficients of the free energy in Eq.\ \eqref{16}.}
\label{barbeta}
\end{figure}

We focus on the quadratic coefficients first, which has two parts of contributions. We have
\begin{align}
\bar\alpha_s=\frac{1}{g_s}-\bar\Pi_s,~~\bar\alpha_p=\frac{1}{g_p}-\bar\Pi_p,~~\alpha_{sp}=\frac{1}{g_{sp}}-\Pi_{sp},
\label{a4}
\end{align}
where we present the diagrams for $\bar\Pi_s$, $\bar\Pi_p$, and $\bar\Pi_{sp}$ in Fig.\ \ref{barbeta}. We will later determine the value of $g_{s,p,sp}$ by matching with Eq.\ \eqref{eq32}.
The polarization operator $\bar\Pi_s$ in the $s$-wave channel is given by
\begin{widetext}
\begin{align}
\bar\Pi_s=&\sum_{m,{\bm k}}\Tr\[ \hat G(\omega_m,\k) (i\sigma^y) \hat G^{T}(-\omega_m,-\k) (i\sigma^y)^\dagger\]
\nonumber\\
=&\sum_{m,{\bm k}}\frac{\Tr\[
\(\begin{array}{cc}
i\omega_m-\epsilon_\k & \lambda\phi e^{-i\theta} \\ \lambda\phi e^{i\theta} & i\omega_m-\epsilon_\k
\end{array}\)
\(\begin{array}{cc}
0 & 1 \\ -1 & 0
\end{array}\)
\(\begin{array}{cc}
-i\omega_m-\epsilon_\k & -\lambda\phi e^{-i\theta} \\ -\lambda\phi e^{i\theta} & -i\omega_m-\epsilon_\k
\end{array}\)^{T}
\(\begin{array}{cc}
0 & 1 \\ -1 & 0
\end{array}\)^\dagger
\]}{[(i\omega_m-\epsilon_\k)^2-\lambda^2\phi ^2][(-i\omega_m-\epsilon_\k)^2-\lambda^2\phi ^2]} \nonumber\\
=&\sum_{m,{\bm k}}\[\frac{1}{(\epsilon_\k+\lambda\phi )^2+\omega_m^2}+\frac{1}{(\epsilon_\k-\lambda\phi )^2+\omega_m^2}\].
\label{50}
\end{align}
The last line intuitively mean that the $s$-wave SC order parameter resides independently on two split FS's. We can do the same for the $p$-wave polarization operator,
\begin{align}
\bar\Pi_p=&\sum_{m,{\bm k}}\Tr\[ \hat G(\omega_m,\k) [\hat\k\cdot (i\bm\sigma \sigma^y)] \hat G^T(-\omega_m,-\k) [\hat\k\cdot (i\bm\sigma \sigma^y)]^\dagger\]
\nonumber\\
=&\sum_{m,{\bm k}}\frac{\Tr\[
\(\begin{array}{cc}
i\omega_m-\epsilon_\k & \lambda\phi e^{-i\theta} \\ \lambda\phi e^{i\theta} & i\omega_m-\epsilon_\k
\end{array}\)
\(\begin{array}{cc}
-e^{-i\theta} & 0 \\ 0 & e^{i\theta}
\end{array}\)
\(\begin{array}{cc}
-i\omega_m-\epsilon_\k & -\lambda\phi e^{-i\theta} \\ -\lambda\phi e^{i\theta} & -i\omega_m-\epsilon_\k
\end{array}\)^T
\(\begin{array}{cc}
-e^{-i\theta} & 0 \\ 0 & e^{i\theta}
\end{array}\)^\dagger
\]}{[(i\omega_m-\epsilon_\k)^2-\lambda^2\phi ^2][(-i\omega_m-\epsilon_\k)^2-\lambda^2\phi ^2]} \nonumber\\
=&\sum_{m,{\bm k}}\[\frac{1}{(\epsilon_\k+\lambda\phi )^2+\omega_m^2}+\frac{1}{(\epsilon_\k-\lambda\phi )^2+\omega_m^2}\],
\label{51}
\end{align}
where we have used the last line of Eq.\ (\ref{sp2}). We note that $\Pi_s=\Pi_p$.\cite{Agterberg-2004} Just like the $s$-wave order parameter, the $p$-wave order parameter independent resides on two split FS's.

Now we move to $\Pi_{sp}$, which is made nonzero by the explicit breaking of inversion symmetry. A straightforward evaluation yields
\begin{align}
\Pi_{sp}=&\sum_{m,{\bm k}}\Tr\[ \hat G(\omega_m,\k) (i\sigma^y) \hat G^T(-\omega_m,-\k) [\hat\k\cdot (i\bm\sigma \sigma^y)]^\dagger\]
\nonumber\\
=&\sum_{m,{\bm k}}\frac{\Tr\[
\(\begin{array}{cc}
i\omega_m-\epsilon_\k & \lambda\phi e^{-i\theta} \\ \lambda\phi e^{i\theta} & i\omega_m-\epsilon_\k
\end{array}\)
\(\begin{array}{cc}
0 & 1 \\ -1 & 0
\end{array}\)
\(\begin{array}{cc}
-i\omega_m-\epsilon_\k & -\lambda\phi e^{-i\theta} \\ -\lambda\phi e^{i\theta} & -i\omega_m-\epsilon_\k
\end{array}\)^T
\(\begin{array}{cc}
-e^{-i\theta} & 0 \\ 0 & e^{i\theta}
\end{array}\)^\dagger
\]}{[(i\omega_m-\epsilon_\k)^2-\lambda^2\phi ^2][(-i\omega_m-\epsilon_\k)^2-\lambda^2\phi ^2]} \nonumber\\
=&\sum_{m,{\bm k}}\[\frac{1}{(\epsilon_\k+\lambda\phi )^2+\omega_m^2}-\frac{1}{(\epsilon_\k-\lambda\phi )^2+\omega_m^2}\].
\label{52}
\end{align}
\end{widetext}

We then see that at quadratic order, the free energy is
\begin{align}
F&=\alpha_s|\Delta_s|^2+\alpha_p|\Delta_p|^2+\alpha_{sp}(\Delta_s\Delta_p^*+\Delta_s^*\Delta_p)+O(\Delta^4)\nonumber\\
&\equiv\alpha_1|\Delta_1|^2+\alpha_2|\Delta_2|^4+\alpha_{12}(\Delta_1\Delta_2^*+\Delta_1^*\Delta_2)+O(\Delta^4),
\end{align}
where we have defined in the second line
\begin{align}
\alpha_1=&\frac{1}{2g_s}+\frac{1}{2g_p}+\frac{1}{2g_{sp}}-\sum_{m,{\bm k}}\frac{1}{(\epsilon_\k+\lambda\phi )^2+\omega_m^2},\nonumber\\
\alpha_2=&\frac{1}{2g_s}+\frac{1}{2g_p}-\frac{1}{2g_{sp}}-\sum_{m,{\bm k}}\frac{1}{(\epsilon_\k-\lambda\phi )^2+\omega_m^2},\nonumber\\
\alpha_{12}=&\frac{1}{2g_s}-\frac{1}{2g_p}.
\end{align}
Comparing with Eq.\ \eqref{eq32} we identify that $\alpha_{12}=\alpha_{C}-\alpha_S$ and that
\begin{align}
\frac{1}{g_s}=&\frac{1}{g_1}+\frac{1}{g_2}+(\alpha_C-\alpha_S),\nonumber\\
\frac{1}{g_p}=&\frac{1}{g_1}+\frac{1}{g_2}-(\alpha_C-\alpha_S),\nonumber\\
\frac{1}{g_{sp}}=&\frac{1}{g_1}-\frac{1}{g_2}.
\label{gsp}
\end{align}
Particularly, we see from the first two lines of Eq.\ \eqref{gsp} that a positive(negative) value of the coefficient $\alpha_C-\alpha_S$ favors $p(s)$-wave SC order, just like we obtained in the main text. 

Next we consider the quartic coeffiecients in Eq.\ (\ref{16}), and we show the relevant diagrams in Fig.\ \ref{barbeta}. The procedure is quite similar to the evaluation of the $\alpha$'s, only in this case one need to evaluate the convolution of four Green functions and four SC vertices. After some lengthy but straightforward calculation, we find that all $\beta$'s can also be expressed in rather simple forms:
\begin{align}
\bar\beta_s&=\bar\beta_p=\frac{1}{4}(\beta_1+\beta_2),~~\bar\beta_m=(\beta_1+\beta_2)\nonumber\\
\bar\beta_m'&=\frac{1}{4}(\beta_1+\beta_2),~~
\beta_{sp}=\beta_{sp}'=\frac{1}{2}(\beta_1-\beta_2),
\end{align}
where
\begin{align}
\beta_{1,2}=\sum_{m,{\bm k}}\frac{1}{[(\epsilon_\k\pm\lambda\phi )^2+\omega_m^2]^2}.\end{align}
are defined the same way as in Eq.\ \eqref{eq33}. 
Plugging these into Eqs.\ \eqref{16}, together with the results at quadratic order, we find that the free energy up to quartic order can be expressed as
\begin{align}
F=&\alpha_1|\Delta_1|^2+\alpha_2|\Delta_2|^2+(\alpha_C-\alpha_S)(\Delta_1\Delta_2^*+\Delta_1^*\Delta_2)\nonumber\\
&+\beta_1|\Delta_1|^4+\beta_2|\Delta_2|^4,
\end{align}
which is identical to Eq.\ \eqref{eq32}.


\begin{thebibliography}{62}%
\makeatletter
\providecommand \@ifxundefined [1]{%
 \@ifx{#1\undefined}
}%
\providecommand \@ifnum [1]{%
 \ifnum #1\expandafter \@firstoftwo
 \else \expandafter \@secondoftwo
 \fi
}%
\providecommand \@ifx [1]{%
 \ifx #1\expandafter \@firstoftwo
 \else \expandafter \@secondoftwo
 \fi
}%
\providecommand \natexlab [1]{#1}%
\providecommand \enquote  [1]{``#1''}%
\providecommand \bibnamefont  [1]{#1}%
\providecommand \bibfnamefont [1]{#1}%
\providecommand \citenamefont [1]{#1}%
\providecommand \href@noop [0]{\@secondoftwo}%
\providecommand \href [0]{\begingroup \@sanitize@url \@href}%
\providecommand \@href[1]{\@@startlink{#1}\@@href}%
\providecommand \@@href[1]{\endgroup#1\@@endlink}%
\providecommand \@sanitize@url [0]{\catcode `\\12\catcode `\$12\catcode
  `\&12\catcode `\#12\catcode `\^12\catcode `\_12\catcode `\%12\relax}%
\providecommand \@@startlink[1]{}%
\providecommand \@@endlink[0]{}%
\providecommand \url  [0]{\begingroup\@sanitize@url \@url }%
\providecommand \@url [1]{\endgroup\@href {#1}{\urlprefix }}%
\providecommand \urlprefix  [0]{URL }%
\providecommand \Eprint [0]{\href }%
\providecommand \doibase [0]{http://dx.doi.org/}%
\providecommand \selectlanguage [0]{\@gobble}%
\providecommand \bibinfo  [0]{\@secondoftwo}%
\providecommand \bibfield  [0]{\@secondoftwo}%
\providecommand \translation [1]{[#1]}%
\providecommand \BibitemOpen [0]{}%
\providecommand \bibitemStop [0]{}%
\providecommand \bibitemNoStop [0]{.\EOS\space}%
\providecommand \EOS [0]{\spacefactor3000\relax}%
\providecommand \BibitemShut  [1]{\csname bibitem#1\endcsname}%
\let\auto@bib@innerbib\@empty
\bibitem [{\citenamefont {{Wen}}(2004)}]{wenbook}%
  \BibitemOpen
  \bibfield  {author} {\bibinfo {author} {\bibfnamefont {X.-G.}\ \bibnamefont
  {{Wen}}},\ }\href@noop {} {\emph {\bibinfo {title} {{Quantum Field Theory of
  Many-body Systems from the Origin of Sound to an Origin of Light and
  Electrons}}}}\ (\bibinfo  {publisher} {Oxford University Press},\ \bibinfo
  {address} {New York},\ \bibinfo {year} {2004})\BibitemShut {NoStop}%
\bibitem [{\citenamefont {Qi}\ and\ \citenamefont {Zhang}(2011)}]{Qi-2010}%
  \BibitemOpen
  \bibfield  {author} {\bibinfo {author} {\bibfnamefont {X.-L.}\ \bibnamefont
  {Qi}}\ and\ \bibinfo {author} {\bibfnamefont {S.-C.}\ \bibnamefont {Zhang}},\
  }\href {\doibase 10.1103/RevModPhys.83.1057} {\bibfield  {journal} {\bibinfo
  {journal} {Rev. Mod. Phys.}\ }\textbf {\bibinfo {volume} {83}},\ \bibinfo
  {pages} {1057} (\bibinfo {year} {2011})}\BibitemShut {NoStop}%
\bibitem [{\citenamefont {Hasan}\ and\ \citenamefont
  {Kane}(2010)}]{Hasan-2010}%
  \BibitemOpen
  \bibfield  {author} {\bibinfo {author} {\bibfnamefont {M.~Z.}\ \bibnamefont
  {Hasan}}\ and\ \bibinfo {author} {\bibfnamefont {C.~L.}\ \bibnamefont
  {Kane}},\ }\href {\doibase 10.1103/RevModPhys.82.3045} {\bibfield  {journal}
  {\bibinfo  {journal} {Rev. Mod. Phys.}\ }\textbf {\bibinfo {volume} {82}},\
  \bibinfo {pages} {3045} (\bibinfo {year} {2010})}\BibitemShut {NoStop}%
\bibitem [{\citenamefont {Qi}\ \emph {et~al.}(2009)\citenamefont {Qi},
  \citenamefont {Hughes}, \citenamefont {Raghu},\ and\ \citenamefont
  {Zhang}}]{qihughesraghuzhang}%
  \BibitemOpen
  \bibfield  {author} {\bibinfo {author} {\bibfnamefont {X.-L.}\ \bibnamefont
  {Qi}}, \bibinfo {author} {\bibfnamefont {T.~L.}\ \bibnamefont {Hughes}},
  \bibinfo {author} {\bibfnamefont {S.}~\bibnamefont {Raghu}}, \ and\ \bibinfo
  {author} {\bibfnamefont {S.-C.}\ \bibnamefont {Zhang}},\ }\href {\doibase
  10.1103/PhysRevLett.102.187001} {\bibfield  {journal} {\bibinfo  {journal}
  {Phys. Rev. Lett.}\ }\textbf {\bibinfo {volume} {102}},\ \bibinfo {pages}
  {187001} (\bibinfo {year} {2009})}\BibitemShut {NoStop}%
\bibitem [{\citenamefont {Schnyder}\ \emph {et~al.}(2008)\citenamefont
  {Schnyder}, \citenamefont {Ryu}, \citenamefont {Furusaki},\ and\
  \citenamefont {Ludwig}}]{Ryu-2008}%
  \BibitemOpen
  \bibfield  {author} {\bibinfo {author} {\bibfnamefont {A.~P.}\ \bibnamefont
  {Schnyder}}, \bibinfo {author} {\bibfnamefont {S.}~\bibnamefont {Ryu}},
  \bibinfo {author} {\bibfnamefont {A.}~\bibnamefont {Furusaki}}, \ and\
  \bibinfo {author} {\bibfnamefont {A.~W.~W.}\ \bibnamefont {Ludwig}},\ }\href
  {\doibase 10.1103/PhysRevB.78.195125} {\bibfield  {journal} {\bibinfo
  {journal} {Phys. Rev. B}\ }\textbf {\bibinfo {volume} {78}},\ \bibinfo
  {pages} {195125} (\bibinfo {year} {2008})}\BibitemShut {NoStop}%
\bibitem [{\citenamefont {Nayak}\ \emph {et~al.}(2008)\citenamefont {Nayak},
  \citenamefont {Simon}, \citenamefont {Stern}, \citenamefont {Freedman},\ and\
  \citenamefont {Das~Sarma}}]{Nayak-2008}%
  \BibitemOpen
  \bibfield  {author} {\bibinfo {author} {\bibfnamefont {C.}~\bibnamefont
  {Nayak}}, \bibinfo {author} {\bibfnamefont {S.~H.}\ \bibnamefont {Simon}},
  \bibinfo {author} {\bibfnamefont {A.}~\bibnamefont {Stern}}, \bibinfo
  {author} {\bibfnamefont {M.}~\bibnamefont {Freedman}}, \ and\ \bibinfo
  {author} {\bibfnamefont {S.}~\bibnamefont {Das~Sarma}},\ }\href {\doibase
  10.1103/RevModPhys.80.1083} {\bibfield  {journal} {\bibinfo  {journal} {Rev.
  Mod. Phys.}\ }\textbf {\bibinfo {volume} {80}},\ \bibinfo {pages} {1083}
  (\bibinfo {year} {2008})}\BibitemShut {NoStop}%
\bibitem [{\citenamefont {Fu}\ and\ \citenamefont {Berg}(2010)}]{Fu-Berg-2010}%
  \BibitemOpen
  \bibfield  {author} {\bibinfo {author} {\bibfnamefont {L.}~\bibnamefont
  {Fu}}\ and\ \bibinfo {author} {\bibfnamefont {E.}~\bibnamefont {Berg}},\
  }\href {\doibase 10.1103/PhysRevLett.105.097001} {\bibfield  {journal}
  {\bibinfo  {journal} {Phys. Rev. Lett.}\ }\textbf {\bibinfo {volume} {105}},\
  \bibinfo {pages} {097001} (\bibinfo {year} {2010})}\BibitemShut {NoStop}%
\bibitem [{\citenamefont {Kriener}\ \emph {et~al.}(2011)\citenamefont
  {Kriener}, \citenamefont {Segawa}, \citenamefont {Ren}, \citenamefont
  {Sasaki},\ and\ \citenamefont {Ando}}]{Kriener-2011}%
  \BibitemOpen
  \bibfield  {author} {\bibinfo {author} {\bibfnamefont {M.}~\bibnamefont
  {Kriener}}, \bibinfo {author} {\bibfnamefont {K.}~\bibnamefont {Segawa}},
  \bibinfo {author} {\bibfnamefont {Z.}~\bibnamefont {Ren}}, \bibinfo {author}
  {\bibfnamefont {S.}~\bibnamefont {Sasaki}}, \ and\ \bibinfo {author}
  {\bibfnamefont {Y.}~\bibnamefont {Ando}},\ }\href {\doibase
  10.1103/PhysRevLett.106.127004} {\bibfield  {journal} {\bibinfo  {journal}
  {Phys. Rev. Lett.}\ }\textbf {\bibinfo {volume} {106}},\ \bibinfo {pages}
  {127004} (\bibinfo {year} {2011})}\BibitemShut {NoStop}%
\bibitem [{\citenamefont {Levy}\ \emph {et~al.}(2013)\citenamefont {Levy},
  \citenamefont {Zhang}, \citenamefont {Ha}, \citenamefont {Sharifi},
  \citenamefont {Talin}, \citenamefont {Kuk},\ and\ \citenamefont
  {Stroscio}}]{Levy-2013}%
  \BibitemOpen
  \bibfield  {author} {\bibinfo {author} {\bibfnamefont {N.}~\bibnamefont
  {Levy}}, \bibinfo {author} {\bibfnamefont {T.}~\bibnamefont {Zhang}},
  \bibinfo {author} {\bibfnamefont {J.}~\bibnamefont {Ha}}, \bibinfo {author}
  {\bibfnamefont {F.}~\bibnamefont {Sharifi}}, \bibinfo {author} {\bibfnamefont
  {A.~A.}\ \bibnamefont {Talin}}, \bibinfo {author} {\bibfnamefont
  {Y.}~\bibnamefont {Kuk}}, \ and\ \bibinfo {author} {\bibfnamefont {J.~A.}\
  \bibnamefont {Stroscio}},\ }\href {\doibase 10.1103/PhysRevLett.110.117001}
  {\bibfield  {journal} {\bibinfo  {journal} {Phys. Rev. Lett.}\ }\textbf
  {\bibinfo {volume} {110}},\ \bibinfo {pages} {117001} (\bibinfo {year}
  {2013})}\BibitemShut {NoStop}%
\bibitem [{\citenamefont {Fu}(2014)}]{Fu-2014}%
  \BibitemOpen
  \bibfield  {author} {\bibinfo {author} {\bibfnamefont {L.}~\bibnamefont
  {Fu}},\ }\href {\doibase 10.1103/PhysRevB.90.100509} {\bibfield  {journal}
  {\bibinfo  {journal} {Phys. Rev. B}\ }\textbf {\bibinfo {volume} {90}},\
  \bibinfo {pages} {100509} (\bibinfo {year} {2014})}\BibitemShut {NoStop}%
\bibitem [{\citenamefont {Wan}\ and\ \citenamefont
  {Savrasov}(2014)}]{Wan-2014}%
  \BibitemOpen
  \bibfield  {author} {\bibinfo {author} {\bibfnamefont {X.}~\bibnamefont
  {Wan}}\ and\ \bibinfo {author} {\bibfnamefont {S.~Y.}\ \bibnamefont
  {Savrasov}},\ }\href {http://dx.doi.org/10.1038/ncomms5144} {\bibfield
  {journal} {\bibinfo  {journal} {Nat Commun}\ }\textbf {\bibinfo {volume} {5}}
  (\bibinfo {year} {2014})}\BibitemShut {NoStop}%
\bibitem [{\citenamefont {Nakosai}\ \emph {et~al.}(2012)\citenamefont
  {Nakosai}, \citenamefont {Tanaka},\ and\ \citenamefont
  {Nagaosa}}]{Nikosai-2012}%
  \BibitemOpen
  \bibfield  {author} {\bibinfo {author} {\bibfnamefont {S.}~\bibnamefont
  {Nakosai}}, \bibinfo {author} {\bibfnamefont {Y.}~\bibnamefont {Tanaka}}, \
  and\ \bibinfo {author} {\bibfnamefont {N.}~\bibnamefont {Nagaosa}},\ }\href
  {\doibase 10.1103/PhysRevLett.108.147003} {\bibfield  {journal} {\bibinfo
  {journal} {Phys. Rev. Lett.}\ }\textbf {\bibinfo {volume} {108}},\ \bibinfo
  {pages} {147003} (\bibinfo {year} {2012})}\BibitemShut {NoStop}%
\bibitem [{\citenamefont {Scheurer}\ and\ \citenamefont
  {Schmalian}(2015)}]{Scheurer-2015}%
  \BibitemOpen
  \bibfield  {author} {\bibinfo {author} {\bibfnamefont {M.~S.}\ \bibnamefont
  {Scheurer}}\ and\ \bibinfo {author} {\bibfnamefont {J.}~\bibnamefont
  {Schmalian}},\ }\href@noop {} {\bibfield  {journal} {\bibinfo  {journal} {Nat
  Commun}\ }\textbf {\bibinfo {volume} {6}} (\bibinfo {year}
  {2015})}\BibitemShut {NoStop}%
\bibitem [{\citenamefont {Hosur}\ \emph {et~al.}(2014)\citenamefont {Hosur},
  \citenamefont {Dai}, \citenamefont {Fang},\ and\ \citenamefont
  {Qi}}]{Hosur-2014}%
  \BibitemOpen
  \bibfield  {author} {\bibinfo {author} {\bibfnamefont {P.}~\bibnamefont
  {Hosur}}, \bibinfo {author} {\bibfnamefont {X.}~\bibnamefont {Dai}}, \bibinfo
  {author} {\bibfnamefont {Z.}~\bibnamefont {Fang}}, \ and\ \bibinfo {author}
  {\bibfnamefont {X.-L.}\ \bibnamefont {Qi}},\ }\href {\doibase
  10.1103/PhysRevB.90.045130} {\bibfield  {journal} {\bibinfo  {journal} {Phys.
  Rev. B}\ }\textbf {\bibinfo {volume} {90}},\ \bibinfo {pages} {045130}
  (\bibinfo {year} {2014})}\BibitemShut {NoStop}%
\bibitem [{\citenamefont {Yuan}\ \emph {et~al.}(2014)\citenamefont {Yuan},
  \citenamefont {Mak},\ and\ \citenamefont {Law}}]{Yuan-2014}%
  \BibitemOpen
  \bibfield  {author} {\bibinfo {author} {\bibfnamefont {N.~F.~Q.}\
  \bibnamefont {Yuan}}, \bibinfo {author} {\bibfnamefont {K.~F.}\ \bibnamefont
  {Mak}}, \ and\ \bibinfo {author} {\bibfnamefont {K.~T.}\ \bibnamefont
  {Law}},\ }\href {\doibase 10.1103/PhysRevLett.113.097001} {\bibfield
  {journal} {\bibinfo  {journal} {Phys. Rev. Lett.}\ }\textbf {\bibinfo
  {volume} {113}},\ \bibinfo {pages} {097001} (\bibinfo {year}
  {2014})}\BibitemShut {NoStop}%
\bibitem [{\citenamefont {Yoshida}\ \emph {et~al.}(2015)\citenamefont
  {Yoshida}, \citenamefont {Sigrist},\ and\ \citenamefont
  {Yanase}}]{Yoshida-2015}%
  \BibitemOpen
  \bibfield  {author} {\bibinfo {author} {\bibfnamefont {T.}~\bibnamefont
  {Yoshida}}, \bibinfo {author} {\bibfnamefont {M.}~\bibnamefont {Sigrist}}, \
  and\ \bibinfo {author} {\bibfnamefont {Y.}~\bibnamefont {Yanase}},\ }\href
  {\doibase 10.1103/PhysRevLett.115.027001} {\bibfield  {journal} {\bibinfo
  {journal} {Phys. Rev. Lett.}\ }\textbf {\bibinfo {volume} {115}},\ \bibinfo
  {pages} {027001} (\bibinfo {year} {2015})}\BibitemShut {NoStop}%
\bibitem [{\citenamefont {Yang}\ \emph {et~al.}(2015)\citenamefont {Yang},
  \citenamefont {Liu}, \citenamefont {Zhang}, \citenamefont {Yao},\ and\
  \citenamefont {Lee}}]{Yang-2015}%
  \BibitemOpen
  \bibfield  {author} {\bibinfo {author} {\bibfnamefont {F.}~\bibnamefont
  {Yang}}, \bibinfo {author} {\bibfnamefont {C.-C.}\ \bibnamefont {Liu}},
  \bibinfo {author} {\bibfnamefont {Y.-Z.}\ \bibnamefont {Zhang}}, \bibinfo
  {author} {\bibfnamefont {Y.}~\bibnamefont {Yao}}, \ and\ \bibinfo {author}
  {\bibfnamefont {D.-H.}\ \bibnamefont {Lee}},\ }\href {\doibase
  10.1103/PhysRevB.91.134514} {\bibfield  {journal} {\bibinfo  {journal} {Phys.
  Rev. B}\ }\textbf {\bibinfo {volume} {91}},\ \bibinfo {pages} {134514}
  (\bibinfo {year} {2015})}\BibitemShut {NoStop}%
\bibitem [{\citenamefont {Ando}\ and\ \citenamefont {Fu}(2015)}]{Ando-2015}%
  \BibitemOpen
  \bibfield  {author} {\bibinfo {author} {\bibfnamefont {Y.}~\bibnamefont
  {Ando}}\ and\ \bibinfo {author} {\bibfnamefont {L.}~\bibnamefont {Fu}},\
  }\href {\doibase 10.1146/annurev-conmatphys-031214-014501} {\bibfield
  {journal} {\bibinfo  {journal} {Ann. Rev. of Cond. Mat. Phys.}\ }\textbf
  {\bibinfo {volume} {6}},\ \bibinfo {pages} {361} (\bibinfo {year}
  {2015})}\BibitemShut {NoStop}%
\bibitem [{\citenamefont {Sato}(2010)}]{Sato-2010}%
  \BibitemOpen
  \bibfield  {author} {\bibinfo {author} {\bibfnamefont {M.}~\bibnamefont
  {Sato}},\ }\href {\doibase 10.1103/PhysRevB.81.220504} {\bibfield  {journal}
  {\bibinfo  {journal} {Phys. Rev. B}\ }\textbf {\bibinfo {volume} {81}},\
  \bibinfo {pages} {220504} (\bibinfo {year} {2010})}\BibitemShut {NoStop}%
\bibitem [{\citenamefont {Leggett}(1975)}]{Leggett-1975}%
  \BibitemOpen
  \bibfield  {author} {\bibinfo {author} {\bibfnamefont {A.~J.}\ \bibnamefont
  {Leggett}},\ }\href {\doibase 10.1103/RevModPhys.47.331} {\bibfield
  {journal} {\bibinfo  {journal} {Rev. Mod. Phys.}\ }\textbf {\bibinfo {volume}
  {47}},\ \bibinfo {pages} {331} (\bibinfo {year} {1975})}\BibitemShut
  {NoStop}%
\bibitem [{\citenamefont {Vollhardt}\ and\ \citenamefont
  {Wolfle}(1990)}]{Vollhardt-1990}%
  \BibitemOpen
  \bibfield  {author} {\bibinfo {author} {\bibfnamefont {D.}~\bibnamefont
  {Vollhardt}}\ and\ \bibinfo {author} {\bibfnamefont {P.}~\bibnamefont
  {Wolfle}},\ }\href@noop {} {\emph {\bibinfo {title} {The superfluid phases of
  Helium 3}}}\ (\bibinfo  {publisher} {Taylor \& Francis},\ \bibinfo {address}
  {London, UK},\ \bibinfo {year} {1990})\BibitemShut {NoStop}%
\bibitem [{\citenamefont {Miyake}\ \emph {et~al.}(1986)\citenamefont {Miyake},
  \citenamefont {Schmitt-Rink},\ and\ \citenamefont {Varma}}]{Varma-1986}%
  \BibitemOpen
  \bibfield  {author} {\bibinfo {author} {\bibfnamefont {K.}~\bibnamefont
  {Miyake}}, \bibinfo {author} {\bibfnamefont {S.}~\bibnamefont
  {Schmitt-Rink}}, \ and\ \bibinfo {author} {\bibfnamefont {C.~M.}\
  \bibnamefont {Varma}},\ }\href {\doibase 10.1103/PhysRevB.34.6554} {\bibfield
   {journal} {\bibinfo  {journal} {Phys. Rev. B}\ }\textbf {\bibinfo {volume}
  {34}},\ \bibinfo {pages} {6554} (\bibinfo {year} {1986})}\BibitemShut
  {NoStop}%
\bibitem [{\citenamefont {Scalapino}(1995)}]{Scalapino-1995}%
  \BibitemOpen
  \bibfield  {author} {\bibinfo {author} {\bibfnamefont {D.}~\bibnamefont
  {Scalapino}},\ }\href {\doibase
  http://dx.doi.org/10.1016/0370-1573(94)00086-I} {\bibfield  {journal}
  {\bibinfo  {journal} {Physics Reports}\ }\textbf {\bibinfo {volume} {250}},\
  \bibinfo {pages} {329 } (\bibinfo {year} {1995})}\BibitemShut {NoStop}%
\bibitem [{\citenamefont {Abanov}\ \emph {et~al.}(2003)\citenamefont {Abanov},
  \citenamefont {Chubukov},\ and\ \citenamefont {Schmalian}}]{Abanov-2003}%
  \BibitemOpen
  \bibfield  {author} {\bibinfo {author} {\bibfnamefont {A.}~\bibnamefont
  {Abanov}}, \bibinfo {author} {\bibfnamefont {A.~V.}\ \bibnamefont
  {Chubukov}}, \ and\ \bibinfo {author} {\bibfnamefont {J.}~\bibnamefont
  {Schmalian}},\ }\href {\doibase 10.1080/0001873021000057123} {\bibfield
  {journal} {\bibinfo  {journal} {Advances in Physics}\ }\textbf {\bibinfo
  {volume} {52}},\ \bibinfo {pages} {119} (\bibinfo {year} {2003})}\BibitemShut
  {NoStop}%
\bibitem [{\citenamefont {Metlitski}\ and\ \citenamefont
  {Sachdev}(2010)}]{Metlitski-2010}%
  \BibitemOpen
  \bibfield  {author} {\bibinfo {author} {\bibfnamefont {M.~A.}\ \bibnamefont
  {Metlitski}}\ and\ \bibinfo {author} {\bibfnamefont {S.}~\bibnamefont
  {Sachdev}},\ }\href {\doibase 10.1103/PhysRevB.82.075128} {\bibfield
  {journal} {\bibinfo  {journal} {Phys. Rev. B}\ }\textbf {\bibinfo {volume}
  {82}},\ \bibinfo {pages} {075128} (\bibinfo {year} {2010})}\BibitemShut
  {NoStop}%
\bibitem [{\citenamefont {Wang}\ and\ \citenamefont
  {Chubukov}(2014)}]{Wang-2014}%
  \BibitemOpen
  \bibfield  {author} {\bibinfo {author} {\bibfnamefont {Y.}~\bibnamefont
  {Wang}}\ and\ \bibinfo {author} {\bibfnamefont {A.~V.}~\bibnamefont
  {Chubukov}},\ }\href {\doibase 10.1103/PhysRevB.90.035149} {\bibfield
  {journal} {\bibinfo  {journal} {Phys. Rev. B}\ }\textbf {\bibinfo {volume}
  {90}},\ \bibinfo {pages} {035149} (\bibinfo {year} {2014})}\BibitemShut
  {NoStop}%
\bibitem [{\citenamefont {Chubukov}(2012)}]{Chubukov-2012}%
  \BibitemOpen
  \bibfield  {author} {\bibinfo {author} {\bibfnamefont {A.~V.}~\bibnamefont
  {Chubukov}},\ }\href {\doibase 10.1146/annurev-conmatphys-020911-125055}
  {\bibfield  {journal} {\bibinfo  {journal} {Ann. Rev. Cond. Mat. Phys.}\
  }\textbf {\bibinfo {volume} {3}},\ \bibinfo {pages} {57} (\bibinfo {year}
  {2012})}\BibitemShut {NoStop}%
\bibitem [{\citenamefont {Kozii}\ and\ \citenamefont {Fu}(2015)}]{Fu-2015}%
  \BibitemOpen
  \bibfield  {author} {\bibinfo {author} {\bibfnamefont {V.}~\bibnamefont
  {Kozii}}\ and\ \bibinfo {author} {\bibfnamefont {L.}~\bibnamefont {Fu}},\
  }\href {\doibase 10.1103/PhysRevLett.115.207002} {\bibfield  {journal}
  {\bibinfo  {journal} {Phys. Rev. Lett.}\ }\textbf {\bibinfo {volume} {115}},\
  \bibinfo {pages} {207002} (\bibinfo {year} {2015})}\BibitemShut {NoStop}%
\bibitem [{\citenamefont {Fu}(2015)}]{Fu-2015a}%
  \BibitemOpen
  \bibfield  {author} {\bibinfo {author} {\bibfnamefont {L.}~\bibnamefont
  {Fu}},\ }\href {\doibase 10.1103/PhysRevLett.115.026401} {\bibfield
  {journal} {\bibinfo  {journal} {Phys. Rev. Lett.}\ }\textbf {\bibinfo
  {volume} {115}},\ \bibinfo {pages} {026401} (\bibinfo {year}
  {2015})}\BibitemShut {NoStop}%
\bibitem [{\citenamefont {Wu}\ and\ \citenamefont {Zhang}(2004)}]{Wu-2004}%
  \BibitemOpen
  \bibfield  {author} {\bibinfo {author} {\bibfnamefont {C.}~\bibnamefont
  {Wu}}\ and\ \bibinfo {author} {\bibfnamefont {S.-C.}\ \bibnamefont {Zhang}},\
  }\href@noop {} {\bibfield  {journal} {\bibinfo  {journal} {Phys. Rev. Lett.}\
  }\textbf {\bibinfo {volume} {93}},\ \bibinfo {pages} {036403} (\bibinfo
  {year} {2004})}\BibitemShut {NoStop}%
\bibitem [{\citenamefont {Wu}\ \emph {et~al.}(2007)\citenamefont {Wu},
  \citenamefont {Sun}, \citenamefont {Fradkin},\ and\ \citenamefont
  {Zhang}}]{Wu-2007}%
  \BibitemOpen
  \bibfield  {author} {\bibinfo {author} {\bibfnamefont {C.~J.}\ \bibnamefont
  {Wu}}, \bibinfo {author} {\bibfnamefont {K.}~\bibnamefont {Sun}}, \bibinfo
  {author} {\bibfnamefont {E.}~\bibnamefont {Fradkin}}, \ and\ \bibinfo
  {author} {\bibfnamefont {S.-C.}\ \bibnamefont {Zhang}},\ }\href@noop {}
  {\bibfield  {journal} {\bibinfo  {journal} {Phys. Rev. B}\ }\textbf {\bibinfo
  {volume} {75}},\ \bibinfo {pages} {115103} (\bibinfo {year}
  {2007})}\BibitemShut {NoStop}%
\bibitem [{\citenamefont {Ezawa}\ \emph {et~al.}(2013)\citenamefont {Ezawa},
  \citenamefont {Tanaka},\ and\ \citenamefont {Nagaosa}}]{Nagaosa-2013}%
  \BibitemOpen
  \bibfield  {author} {\bibinfo {author} {\bibfnamefont {M.}~\bibnamefont
  {Ezawa}}, \bibinfo {author} {\bibfnamefont {Y.}~\bibnamefont {Tanaka}}, \
  and\ \bibinfo {author} {\bibfnamefont {N.}~\bibnamefont {Nagaosa}},\ }\href
  {http://dx.doi.org/10.1038/srep02790} {\bibfield  {journal} {\bibinfo
  {journal} {Scientific Reports}\ }\textbf {\bibinfo {volume} {3}},\ \bibinfo
  {pages} {2790 EP } (\bibinfo {year} {2013})}\BibitemShut {NoStop}%
\bibitem [{\citenamefont {Babaev}(2004)}]{babaev}%
  \BibitemOpen
  \bibfield  {author} {\bibinfo {author} {\bibfnamefont {E.}~\bibnamefont
  {Babaev}},\ }\href@noop {} {\bibfield  {journal} {\bibinfo  {journal} {Nucl.
  Phys. B}\ }\textbf {\bibinfo {volume} {686}},\ \bibinfo {pages} {397}
  (\bibinfo {year} {2004})}\BibitemShut {NoStop}%
\bibitem [{\citenamefont {Berg}\ \emph
  {et~al.}(2009{\natexlab{a}})\citenamefont {Berg}, \citenamefont {Fradkin},\
  and\ \citenamefont {Kivelson}}]{Berg-2009}%
  \BibitemOpen
  \bibfield  {author} {\bibinfo {author} {\bibfnamefont {E.}~\bibnamefont
  {Berg}}, \bibinfo {author} {\bibfnamefont {E.}~\bibnamefont {Fradkin}}, \
  and\ \bibinfo {author} {\bibfnamefont {S.~A.}\ \bibnamefont {Kivelson}},\
  }\href@noop {} {\bibfield  {journal} {\bibinfo  {journal} {Nat. Phys.}\
  }\textbf {\bibinfo {volume} {5}},\ \bibinfo {pages} {830} (\bibinfo {year}
  {2009}{\natexlab{a}})}\BibitemShut {NoStop}%
\bibitem [{\citenamefont {Read}\ and\ \citenamefont {Green}(2000)}]{readgreen}%
  \BibitemOpen
  \bibfield  {author} {\bibinfo {author} {\bibfnamefont {N.}~\bibnamefont
  {Read}}\ and\ \bibinfo {author} {\bibfnamefont {D.}~\bibnamefont {Green}},\
  }\href {\doibase 10.1103/PhysRevB.61.10267} {\bibfield  {journal} {\bibinfo
  {journal} {Phys. Rev. B}\ }\textbf {\bibinfo {volume} {61}},\ \bibinfo
  {pages} {10267} (\bibinfo {year} {2000})}\BibitemShut {NoStop}%
\bibitem [{\citenamefont {Ivanov}(2001)}]{ivanov}%
  \BibitemOpen
  \bibfield  {author} {\bibinfo {author} {\bibfnamefont {D.~A.}\ \bibnamefont
  {Ivanov}},\ }\href {\doibase 10.1103/PhysRevLett.86.268} {\bibfield
  {journal} {\bibinfo  {journal} {Phys. Rev. Lett.}\ }\textbf {\bibinfo
  {volume} {86}},\ \bibinfo {pages} {268} (\bibinfo {year} {2001})}\BibitemShut
  {NoStop}%
\bibitem [{\citenamefont {Rice}\ and\ \citenamefont
  {Sigrist}(1996)}]{Rice-1996}%
  \BibitemOpen
  \bibfield  {author} {\bibinfo {author} {\bibfnamefont {T.~M.}\ \bibnamefont
  {Rice}}\ and\ \bibinfo {author} {\bibfnamefont {M.}~\bibnamefont {Sigrist}},\
  }\href@noop {} {\bibfield  {journal} {\bibinfo  {journal} {J. Phys. C}\
  }\textbf {\bibinfo {volume} {7}},\ \bibinfo {pages} {L643} (\bibinfo {year}
  {1996})}\BibitemShut {NoStop}%
\bibitem [{\citenamefont {Raghu}\ \emph {et~al.}(2010)\citenamefont {Raghu},
  \citenamefont {Kapitulnik},\ and\ \citenamefont {Kivelson}}]{Raghu-2010}%
  \BibitemOpen
  \bibfield  {author} {\bibinfo {author} {\bibfnamefont {S.}~\bibnamefont
  {Raghu}}, \bibinfo {author} {\bibfnamefont {A.}~\bibnamefont {Kapitulnik}}, \
  and\ \bibinfo {author} {\bibfnamefont {S.~A.}\ \bibnamefont {Kivelson}},\
  }\href {\doibase 10.1103/PhysRevLett.105.136401} {\bibfield  {journal}
  {\bibinfo  {journal} {Phys. Rev. Lett.}\ }\textbf {\bibinfo {volume} {105}},\
  \bibinfo {pages} {136401} (\bibinfo {year} {2010})}\BibitemShut {NoStop}%
\bibitem [{\citenamefont {Millis}\ \emph {et~al.}(1988)\citenamefont {Millis},
  \citenamefont {Sachdev},\ and\ \citenamefont {Varma}}]{Millis-1988}%
  \BibitemOpen
  \bibfield  {author} {\bibinfo {author} {\bibfnamefont {A.~J.}\ \bibnamefont
  {Millis}}, \bibinfo {author} {\bibfnamefont {S.}~\bibnamefont {Sachdev}}, \
  and\ \bibinfo {author} {\bibfnamefont {C.~M.}\ \bibnamefont {Varma}},\ }\href
  {\doibase 10.1103/PhysRevB.37.4975} {\bibfield  {journal} {\bibinfo
  {journal} {Phys. Rev. B}\ }\textbf {\bibinfo {volume} {37}},\ \bibinfo
  {pages} {4975} (\bibinfo {year} {1988})}\BibitemShut {NoStop}%
\bibitem [{\citenamefont {Abanov}\ \emph {et~al.}(2001)\citenamefont {Abanov},
  \citenamefont {Chubukov},\ and\ \citenamefont {Finkel'stein}}]{acf-2001}%
  \BibitemOpen
  \bibfield  {author} {\bibinfo {author} {\bibfnamefont {A.}~\bibnamefont
  {Abanov}}, \bibinfo {author} {\bibfnamefont {A.~V.}\ \bibnamefont
  {Chubukov}}, \ and\ \bibinfo {author} {\bibfnamefont {A.~M.}\ \bibnamefont
  {Finkel'stein}},\ }\href {http://stacks.iop.org/0295-5075/54/i=4/a=488}
  {\bibfield  {journal} {\bibinfo  {journal} {EPL (Europhysics Letters)}\
  }\textbf {\bibinfo {volume} {54}},\ \bibinfo {pages} {488} (\bibinfo {year}
  {2001})}\BibitemShut {NoStop}%
\bibitem [{\citenamefont {Efetov}\ \emph {et~al.}(2013)\citenamefont {Efetov},
  \citenamefont {Meier},\ and\ \citenamefont {Pepin}}]{Efetov-2013}%
  \BibitemOpen
  \bibfield  {author} {\bibinfo {author} {\bibfnamefont {K.~B.}\ \bibnamefont
  {Efetov}}, \bibinfo {author} {\bibfnamefont {H.}~\bibnamefont {Meier}}, \
  and\ \bibinfo {author} {\bibfnamefont {C.}~\bibnamefont {Pepin}},\ }\href
  {http://dx.doi.org/10.1038/nphys2641} {\bibfield  {journal} {\bibinfo
  {journal} {Nat Phys}\ }\textbf {\bibinfo {volume} {9}},\ \bibinfo {pages}
  {442} (\bibinfo {year} {2013})}\BibitemShut {NoStop}%
\bibitem [{\citenamefont {Wang}\ and\ \citenamefont
  {Chubukov}(2013{\natexlab{a}})}]{Wang-2013}%
  \BibitemOpen
  \bibfield  {author} {\bibinfo {author} {\bibfnamefont {Y.}~\bibnamefont
  {Wang}}\ and\ \bibinfo {author} {\bibfnamefont {A.~V.}\ \bibnamefont
  {Chubukov}},\ }\href {\doibase 10.1103/PhysRevLett.110.127001} {\bibfield
  {journal} {\bibinfo  {journal} {Phys. Rev. Lett.}\ }\textbf {\bibinfo
  {volume} {110}},\ \bibinfo {pages} {127001} (\bibinfo {year}
  {2013}{\natexlab{a}})}\BibitemShut {NoStop}%
\bibitem [{\citenamefont {Metlitski}\ \emph {et~al.}(2015)\citenamefont
  {Metlitski}, \citenamefont {Mross}, \citenamefont {Sachdev},\ and\
  \citenamefont {Senthil}}]{Metlitski-2015}%
  \BibitemOpen
  \bibfield  {author} {\bibinfo {author} {\bibfnamefont {M.~A.}\ \bibnamefont
  {Metlitski}}, \bibinfo {author} {\bibfnamefont {D.~F.}\ \bibnamefont
  {Mross}}, \bibinfo {author} {\bibfnamefont {S.}~\bibnamefont {Sachdev}}, \
  and\ \bibinfo {author} {\bibfnamefont {T.}~\bibnamefont {Senthil}},\ }\href
  {\doibase 10.1103/PhysRevB.91.115111} {\bibfield  {journal} {\bibinfo
  {journal} {Phys. Rev. B}\ }\textbf {\bibinfo {volume} {91}},\ \bibinfo
  {pages} {115111} (\bibinfo {year} {2015})}\BibitemShut {NoStop}%
\bibitem [{\citenamefont {{Schattner}}\ \emph {et~al.}(2015)\citenamefont
  {{Schattner}}, \citenamefont {{Lederer}}, \citenamefont {{Kivelson}},\ and\
  \citenamefont {{Berg}}}]{Schattner-2015}%
  \BibitemOpen
  \bibfield  {author} {\bibinfo {author} {\bibfnamefont {Y.}~\bibnamefont
  {{Schattner}}}, \bibinfo {author} {\bibfnamefont {S.}~\bibnamefont
  {{Lederer}}}, \bibinfo {author} {\bibfnamefont {S.~A.}\ \bibnamefont
  {{Kivelson}}}, \ and\ \bibinfo {author} {\bibfnamefont {E.}~\bibnamefont
  {{Berg}}},\ }\href@noop {} {\enquote {\bibinfo {title} {{Ising nematic
  quantum critical point in a metal: a Monte Carlo study}},}\ } (\bibinfo
  {year} {2015}),\ \Eprint {http://arxiv.org/abs/arXiv:1511.03282}
  {arXiv:1511.03282} \BibitemShut {NoStop}%
\bibitem [{\citenamefont {Raghu}\ \emph {et~al.}(2015)\citenamefont {Raghu},
  \citenamefont {Torroba},\ and\ \citenamefont {Wang}}]{Raghu-2015}%
  \BibitemOpen
  \bibfield  {author} {\bibinfo {author} {\bibfnamefont {S.}~\bibnamefont
  {Raghu}}, \bibinfo {author} {\bibfnamefont {G.}~\bibnamefont {Torroba}}, \
  and\ \bibinfo {author} {\bibfnamefont {H.}~\bibnamefont {Wang}},\ }\href
  {\doibase 10.1103/PhysRevB.92.205104} {\bibfield  {journal} {\bibinfo
  {journal} {Phys. Rev. B}\ }\textbf {\bibinfo {volume} {92}},\ \bibinfo
  {pages} {205104} (\bibinfo {year} {2015})}\BibitemShut {NoStop}%
\bibitem [{\citenamefont {Wang}\ and\ \citenamefont
  {Chubukov}(2013{\natexlab{b}})}]{Wang-2013a}%
  \BibitemOpen
  \bibfield  {author} {\bibinfo {author} {\bibfnamefont {Y.}~\bibnamefont
  {Wang}}\ and\ \bibinfo {author} {\bibfnamefont {A.~V.}~\bibnamefont
  {Chubukov}},\ }\href {\doibase 10.1103/PhysRevB.88.024516} {\bibfield
  {journal} {\bibinfo  {journal} {Phys. Rev. B}\ }\textbf {\bibinfo {volume}
  {88}},\ \bibinfo {pages} {024516} (\bibinfo {year}
  {2013}{\natexlab{b}})}\BibitemShut {NoStop}%
\bibitem [{\citenamefont {Wang}\ and\ \citenamefont
  {Chubukov}(2015{\natexlab{a}})}]{Wang-2015}%
  \BibitemOpen
  \bibfield  {author} {\bibinfo {author} {\bibfnamefont {Y.}~\bibnamefont
  {Wang}}\ and\ \bibinfo {author} {\bibfnamefont {A.~V.}~\bibnamefont
  {Chubukov}},\ }\href {\doibase 10.1103/PhysRevB.91.195113} {\bibfield
  {journal} {\bibinfo  {journal} {Phys. Rev. B}\ }\textbf {\bibinfo {volume}
  {91}},\ \bibinfo {pages} {195113} (\bibinfo {year}
  {2015}{\natexlab{a}})}\BibitemShut {NoStop}%
\bibitem [{\citenamefont {Wang}\ and\ \citenamefont
  {Chubukov}(2015{\natexlab{b}})}]{Wang-2015a}%
  \BibitemOpen
  \bibfield  {author} {\bibinfo {author} {\bibfnamefont {Y.}~\bibnamefont
  {Wang}}\ and\ \bibinfo {author} {\bibfnamefont {A.~V.}\ \bibnamefont
  {Chubukov}},\ }\href {\doibase 10.1103/PhysRevB.92.125108} {\bibfield
  {journal} {\bibinfo  {journal} {Phys. Rev. B}\ }\textbf {\bibinfo {volume}
  {92}},\ \bibinfo {pages} {125108} (\bibinfo {year}
  {2015}{\natexlab{b}})}\BibitemShut {NoStop}%
\bibitem [{\citenamefont {Frigeri}\ \emph {et~al.}(2004)\citenamefont
  {Frigeri}, \citenamefont {Agterberg}, \citenamefont {Koga},\ and\
  \citenamefont {Sigrist}}]{Agterberg-2004}%
  \BibitemOpen
  \bibfield  {author} {\bibinfo {author} {\bibfnamefont {P.~A.}\ \bibnamefont
  {Frigeri}}, \bibinfo {author} {\bibfnamefont {D.~F.}\ \bibnamefont
  {Agterberg}}, \bibinfo {author} {\bibfnamefont {A.}~\bibnamefont {Koga}}, \
  and\ \bibinfo {author} {\bibfnamefont {M.}~\bibnamefont {Sigrist}},\ }\href
  {\doibase 10.1103/PhysRevLett.92.097001} {\bibfield  {journal} {\bibinfo
  {journal} {Phys. Rev. Lett.}\ }\textbf {\bibinfo {volume} {92}},\ \bibinfo
  {pages} {097001} (\bibinfo {year} {2004})}\BibitemShut {NoStop}%
\bibitem [{\citenamefont {Qi}\ \emph {et~al.}(2010)\citenamefont {Qi},
  \citenamefont {Hughes},\ and\ \citenamefont {Zhang}}]{Qi-2010a}%
  \BibitemOpen
  \bibfield  {author} {\bibinfo {author} {\bibfnamefont {X.-L.}\ \bibnamefont
  {Qi}}, \bibinfo {author} {\bibfnamefont {T.~L.}\ \bibnamefont {Hughes}}, \
  and\ \bibinfo {author} {\bibfnamefont {S.-C.}\ \bibnamefont {Zhang}},\ }\href
  {\doibase 10.1103/PhysRevB.81.134508} {\bibfield  {journal} {\bibinfo
  {journal} {Phys. Rev. B}\ }\textbf {\bibinfo {volume} {81}},\ \bibinfo
  {pages} {134508} (\bibinfo {year} {2010})}\BibitemShut {NoStop}%
\bibitem [{\citenamefont {Volovik}(1999)}]{Volovik_1999}%
  \BibitemOpen
  \bibfield  {author} {\bibinfo {author} {\bibfnamefont {G.~E.}\ \bibnamefont
  {Volovik}},\ }\href {\doibase 10.1134/1.568231} {\bibfield  {journal}
  {\bibinfo  {journal} {Journal of Experimental and Theoretical Physics
  Letters}\ }\textbf {\bibinfo {volume} {70}},\ \bibinfo {pages} {792}
  (\bibinfo {year} {1999})}\BibitemShut {NoStop}%
\bibitem [{\citenamefont {Berg}\ \emph
  {et~al.}(2009{\natexlab{b}})\citenamefont {Berg}, \citenamefont {Fradkin},\
  and\ \citenamefont {Kivelson}}]{Berg-2009b}%
  \BibitemOpen
  \bibfield  {author} {\bibinfo {author} {\bibfnamefont {E.}~\bibnamefont
  {Berg}}, \bibinfo {author} {\bibfnamefont {E.}~\bibnamefont {Fradkin}}, \
  and\ \bibinfo {author} {\bibfnamefont {S.~A.}\ \bibnamefont {Kivelson}},\
  }\href {\doibase 10.1103/PhysRevB.79.064515} {\bibfield  {journal} {\bibinfo
  {journal} {Phys. Rev. B}\ }\textbf {\bibinfo {volume} {79}},\ \bibinfo
  {pages} {064515} (\bibinfo {year} {2009}{\natexlab{b}})}\BibitemShut
  {NoStop}%
\bibitem [{\citenamefont {Agterberg}\ and\ \citenamefont
  {Tsunetsugu}(2008)}]{Agterberg-2008}%
  \BibitemOpen
  \bibfield  {author} {\bibinfo {author} {\bibfnamefont {D.~F.}\ \bibnamefont
  {Agterberg}}\ and\ \bibinfo {author} {\bibfnamefont {H.}~\bibnamefont
  {Tsunetsugu}},\ }\href@noop {} {\bibfield  {journal} {\bibinfo  {journal}
  {Nat. Phys.}\ }\textbf {\bibinfo {volume} {4}},\ \bibinfo {pages} {639}
  (\bibinfo {year} {2008})}\BibitemShut {NoStop}%
\bibitem [{\citenamefont {Vakaryuk}\ and\ \citenamefont
  {Leggett}(2009)}]{Vakaryuk-2009}%
  \BibitemOpen
  \bibfield  {author} {\bibinfo {author} {\bibfnamefont {V.}~\bibnamefont
  {Vakaryuk}}\ and\ \bibinfo {author} {\bibfnamefont {A.~J.}\ \bibnamefont
  {Leggett}},\ }\href {\doibase 10.1103/PhysRevLett.103.057003} {\bibfield
  {journal} {\bibinfo  {journal} {Phys. Rev. Lett.}\ }\textbf {\bibinfo
  {volume} {103}},\ \bibinfo {pages} {057003} (\bibinfo {year}
  {2009})}\BibitemShut {NoStop}%
\bibitem [{\citenamefont {Fu}\ and\ \citenamefont {Kane}(2008)}]{Fu-2008}%
  \BibitemOpen
  \bibfield  {author} {\bibinfo {author} {\bibfnamefont {L.}~\bibnamefont
  {Fu}}\ and\ \bibinfo {author} {\bibfnamefont {C.~L.}\ \bibnamefont {Kane}},\
  }\href {\doibase 10.1103/PhysRevLett.100.096407} {\bibfield  {journal}
  {\bibinfo  {journal} {Phys. Rev. Lett.}\ }\textbf {\bibinfo {volume} {100}},\
  \bibinfo {pages} {096407} (\bibinfo {year} {2008})}\BibitemShut {NoStop}%
\bibitem [{\citenamefont {Roy}(2010)}]{Roy-2010}%
  \BibitemOpen
  \bibfield  {author} {\bibinfo {author} {\bibfnamefont {R.}~\bibnamefont
  {Roy}},\ }\href {\doibase 10.1103/PhysRevLett.105.186401} {\bibfield
  {journal} {\bibinfo  {journal} {Phys. Rev. Lett.}\ }\textbf {\bibinfo
  {volume} {105}},\ \bibinfo {pages} {186401} (\bibinfo {year}
  {2010})}\BibitemShut {NoStop}%
\bibitem [{\citenamefont {Babaev}\ \emph {et~al.}(2001)\citenamefont {Babaev},
  \citenamefont {Faddeev},\ and\ \citenamefont {Niemi}}]{babaev2}%
  \BibitemOpen
  \bibfield  {author} {\bibinfo {author} {\bibfnamefont {E.}~\bibnamefont
  {Babaev}}, \bibinfo {author} {\bibfnamefont {L.~D.}\ \bibnamefont {Faddeev}},
  \ and\ \bibinfo {author} {\bibfnamefont {A.~J.}\ \bibnamefont {Niemi}},\
  }\href@noop {} {\bibfield  {journal} {\bibinfo  {journal} {Phys. Rev. B}\
  }\textbf {\bibinfo {volume} {65}},\ \bibinfo {pages} {100512(R)} (\bibinfo
  {year} {2001})}\BibitemShut {NoStop}%
\bibitem [{\citenamefont {Faddeev}\ and\ \citenamefont
  {Niemi}(1997)}]{faddeev}%
  \BibitemOpen
  \bibfield  {author} {\bibinfo {author} {\bibfnamefont {L.~D.}\ \bibnamefont
  {Faddeev}}\ and\ \bibinfo {author} {\bibfnamefont {A.~J.}\ \bibnamefont
  {Niemi}},\ }\href@noop {} {\bibfield  {journal} {\bibinfo  {journal}
  {Nature}\ }\textbf {\bibinfo {volume} {387}},\ \bibinfo {pages} {58}
  (\bibinfo {year} {1997})}\BibitemShut {NoStop}%
\bibitem [{\citenamefont {Chung}\ \emph {et~al.}(2013)\citenamefont {Chung},
  \citenamefont {Horowitz},\ and\ \citenamefont {Qi}}]{Chung-2013}%
  \BibitemOpen
  \bibfield  {author} {\bibinfo {author} {\bibfnamefont {S.~B.}\ \bibnamefont
  {Chung}}, \bibinfo {author} {\bibfnamefont {J.}~\bibnamefont {Horowitz}}, \
  and\ \bibinfo {author} {\bibfnamefont {X.-L.}\ \bibnamefont {Qi}},\ }\href
  {http://link.aps.org/doi/10.1103/PhysRevB.88.214514} {\bibfield  {journal}
  {\bibinfo  {journal} {Physical Review B}\ }\textbf {\bibinfo {volume} {88}},\
  \bibinfo {pages} {214514} (\bibinfo {year} {2013})}\BibitemShut {NoStop}%
\bibitem [{\citenamefont {Leggett}(1966)}]{Leggett-1966}%
  \BibitemOpen
  \bibfield  {author} {\bibinfo {author} {\bibfnamefont {A.~J.}\ \bibnamefont
  {Leggett}},\ }\href@noop {} {\bibfield  {journal} {\bibinfo  {journal} {Prog.
  Theor. Phys.}\ }\textbf {\bibinfo {volume} {36}},\ \bibinfo {pages} {901}
  (\bibinfo {year} {1966})}\BibitemShut {NoStop}%
\bibitem [{\citenamefont {Klein}(2010)}]{Klein-2010}%
  \BibitemOpen
  \bibfield  {author} {\bibinfo {author} {\bibfnamefont {M.~V.}\ \bibnamefont
  {Klein}},\ }\href {\doibase 10.1103/PhysRevB.82.014507} {\bibfield  {journal}
  {\bibinfo  {journal} {Phys. Rev. B}\ }\textbf {\bibinfo {volume} {82}},\
  \bibinfo {pages} {014507} (\bibinfo {year} {2010})}\BibitemShut {NoStop}%
\bibitem [{\citenamefont {{Venderbos}}\ \emph {et~al.}(2015)\citenamefont
  {{Venderbos}}, \citenamefont {{Kozii}},\ and\ \citenamefont
  {{Fu}}}]{Fu-2015b}%
  \BibitemOpen
  \bibfield  {author} {\bibinfo {author} {\bibfnamefont {J.~W.~F.}\
  \bibnamefont {{Venderbos}}}, \bibinfo {author} {\bibfnamefont
  {V.}~\bibnamefont {{Kozii}}}, \ and\ \bibinfo {author} {\bibfnamefont
  {L.}~\bibnamefont {{Fu}}},\ }\href@noop {} {\bibfield  {journal} {\bibinfo
  {journal} {ArXiv e-prints}\ } (\bibinfo {year} {2015})},\ \Eprint
  {http://arxiv.org/abs/1512.04554} {arXiv:1512.04554 [cond-mat.supr-con]}
  \BibitemShut {NoStop}%
\end{thebibliography}
\end{document}